\documentclass[11pt,a4paper]{article}

\usepackage[T1]{fontenc}
\usepackage{graphicx}
\usepackage{xcolor}
\usepackage{geometry}
\usepackage{amsmath}
\usepackage{amssymb}
\usepackage{booktabs}
\IfFileExists{enumitem.sty}{%
  \usepackage{enumitem}
  \setlist[itemize]{itemsep=0.25em, leftmargin=1.6em}
  \setlist[enumerate]{itemsep=0.25em, leftmargin=1.8em}
}{}
\usepackage{caption}
\IfFileExists{microtype.sty}{\usepackage{microtype}}{}

\usepackage[numbers,sort&compress]{natbib}
\newcommand{\Rx}{R_x}
\newcommand{\Ry}{R_y}
\newcommand{\Rz}{R_z}
\newcommand{\degC}{$^\circ$C}
\newcommand{\NI}{N_I}
\newcommand{\NO}{N_O}
\newcommand{\NL}{N_L}

\definecolor{darkgray}{gray}{0.25}
\usepackage[
    colorlinks=true,
    linkcolor=blue,
    citecolor=darkgray,
    urlcolor=blue,
    bookmarks=true,
    bookmarksnumbered=true,
    unicode=true
]{hyperref}
\hypersetup{
    pdftitle={Application of a Quantum Machine Learning Algorithm to the Applied Problem of Modeling the Thermal Regime of Servers},
    pdfauthor={P.V. Buzin, M.A. Remnev},
    pdfsubject={Quantum machine learning for predictive maintenance of HPC servers},
    pdfkeywords={QML, quantum machine learning, practical task, prediction analytics}
}

\begin{document}

\title{\textbf{Application a quantum machine learning algorithm to the applied problem of modeling the thermal regime of a servers}}

\author{%
P.V.~Buzin
\and
M.A.~Remnev \thanks{e-mail: \href{mailto:m.remnev@yandex.ru}{m.remnev@yandex.ru}}\\[0.8em]
}

\date{}

\maketitle

\begin{abstract}
\noindent
Quantum computing is a rapidly developed field of science. In the NISC era, quantum machine learning stands out, potentially outperforming classical machine learning. In this paper, we explore the applicability of quantum machine learning to a real-world application. We solved the problem of prediction of the temperature of GPUs in a high-performance computing cluster. We trained a quantum circuit to predict the temperature of each of the eight GPU of each cluster node based on 16 physical parameters. We demonstrated that quantum machine learning performs on this task just as well as classical machine learning. Furthermore, using model problems, we demonstrated that quantum machine learning can potentially outperform classical machine learning.
\end{abstract}

\noindent\textbf{\textit{Keywords:}} QML, quantum machine learning, practical task, prediction analytics

\bigskip

\section{Introduction}
\label{sec:intro}

In the modern era of digital transformation, predictive analytics is becoming a key tool for improving operational efficiency across various industries~\cite{kumar2018,desai2024}. One of the most important practical applications in this area is predictive maintenance, which enables the prediction of equipment failures and optimization of repair schedules, significantly reducing downtime and operating costs. The implementation of predictive analytics enables more precise prediction of equipment failures, thereby reducing operating costs and downtime~\cite{patel2023}.

Machine learning (ML) methods, including deep learning methods, are traditionally used to solve predictive maintenance problems~\cite{li2025}. These methods analyze historical data from equipment sensors, identify hidden patterns, and generate predictions of probable failures. However, with increasing data complexity and dimensionality, traditional ML methods encounter computational limitations, stimulating the search for more powerful computing paradigms~\cite{thompson2023}.

Concurrent with the development of classical ML, rapid progress is occurring in quantum computing --- a technology based on principles of quantum mechanics such as superposition and entanglement~\cite{mitarai2018}. The use of qubits (quantum bits) instead of classical bits opens up the possibility of exponential growth in computing power for certain classes of problems.

Quantum Machine Learning (QML) is a rapidly developing interdisciplinary field at the intersection of quantum computing and machine learning~\cite{peral2024,desdentado2025,schnabel2025}. QML explores how quantum computers can improve or accelerate traditional ML algorithms. The underlying hypothesis is that quantum computers are capable of working with much more complex and high-dimensional feature spaces due to their fundamental properties~\cite{schnabel2025}. Access to these extended Hilbert spaces allows to create the kernel functions that are extremely difficult or impossible to reproduce on classical machines, opening the way to solving problems previously considered computationally intractable~\cite{biamonte2017}.

In recent years, a number of scientific papers have appeared demonstrating the potential effectiveness of qubits in machine learning. Useful properties include nonlinearity~\cite{schuld2019,peral2024}, the use of gradient descent and gradient calculation using the parameter-shift rule~\cite{peral2024,schuld2020}, and the demonstration of good results on simple problems~\cite{ren2022,tolstobrov2024hybrid,li2023}. Furthermore, such factors as the exponential growth of the dimension of the Hilbert space, quantum entanglement, and the unitarity of algorithms support the effectiveness of quantum machine learning.

Experiments on benchmark datasets (such as MNIST datasets) show that hybrid models can achieve competitive or even superior accuracy compared to fully classical machine learning algorithms while simultaneously reducing the number of parameters~\cite{tolstobrov2024hybrid,tolstobrov2024integrated,xiao2023,senokosov2024}. This supports the thesis that quantum circuits are capable of implementing highly expressive and nonlinear transformations that effective for classification and regression problems. Moreover even one qubit can tackle the challenge of classification and regression~\cite{dutta2022,perez2021}.

Despite encouraging results, a significant drawback of the most existing works in QML area are focus on simplified model problems (toy problems) and benchmark datasets of limited size. Typically, demonstrations of a ``quantum advantage'' are limited to general statements about the fundamental properties of quantum systems---superposition, entanglement, and Hilbert space operation---without a deep investigation of the nature of this advantage or the identification of specific types of problems or data dependencies where QML is most effective~\cite{wang2024,zaman2023}. The lack of systematic research aimed at identifying such dependencies, as well as the application of QML to real industrial problems with large volumes of data (big data), hinders the transition from laboratory experiments to field implementation.

The aim of this study is to apply quantum machine learning to a practical predictive maintenance problem using a large dataset of real equipment telemetry data, as well as to systematically investigate the potential advantages of QML over classical neural networks.

To achieve this goal we completed the following set of tasks:
\begin{enumerate}
    \item Conducting a series of experiments with parameterized quantum circuits on regression problems using model functions to identify the types of dependencies and data characteristics for which QML consistently demonstrates better accuracy than classical neural networks.
    \item Selection of the optimal quantum circuit architecture based on a competetive analysis, with demonstrating of the highest efficiency for the identified classes of dependencies.
    \item Testing of the best selected QML architecture on a real-world predictive maintenance problem with a large dataset containing industrial equipment telemetry data.
\end{enumerate}

This two-step approach allows us not only to propose a potentially more efficient solution for a specific application problem, but also to contribute to the fundamental understanding of the applicability of quantum machine learning by moving from general statements to quantitatively substantiated conclusions. We used simulator of quantum computer for modeling quantum circuit. Calculations are carried out without taking into account the noise.

\section{Predictive Maintenance}
\label{sec:predictive}

Predictive analytics, as part of the predictive maintenance process, identifies the increasing risk of equipment failure or failure of specific modules based on real-world data. Predictive analytics enables maintenance and preventative repairs based on predictions of the equipment's condition and remaining service life, rather than following a pre-planned schedule or after a failure.

Predictive analytics includes the following operations:
\begin{itemize}
    \item collecting data from sensors (temperature, vibration, pressure, voltage, current, etc.), collecting telemetry and operation logs, monitoring operating parameters (external factors and their time of action - external temperature, humidity, overload, etc.);
    \item transmitting and storing data for subsequent analysis and statistics collection;
    \item modeling and analytics using statistical methods, machine learning, time series analysis, anomaly detection, factor influence ranking, estimating mean time between failures, and predicting remaining service life;
    \item solving an optimization problem for deciding whether to perform maintenance, taking into account the duration, scope, and cost of maintenance under various scenarios, service availability at specific locations and times, and other factors;
    \item making management decisions about unscheduled maintenance and performing procedures for transferring equipment from operation to maintenance mode.
\end{itemize}

It is important to note that predictive analytics tasks do not replace, but rather complement, operational monitoring tasks.

\section{Task Description. Dataset}
\label{sec:task}

This work addressed the problem of predictive analytics for high-performance servers with GPU that used in HPC clusters for parallel computations.

The node servers are the DGX A100 servers from NVidia, which includes:
\begin{itemize}
    \item 8x NVIDIA A100 Tensor Core GPUs with 640 GB Total GPU Memory,
    \item 2x 64-Core AMD Rome CPUs,
    \item 2 TB RAM,
    \item SSD,
    \item NVSwitches and Network Interfaces,
    \item 6x Power Supply Units (PSU).
\end{itemize}

\begin{figure}[htbp]
    \centering
    \includegraphics[width=0.75\linewidth]{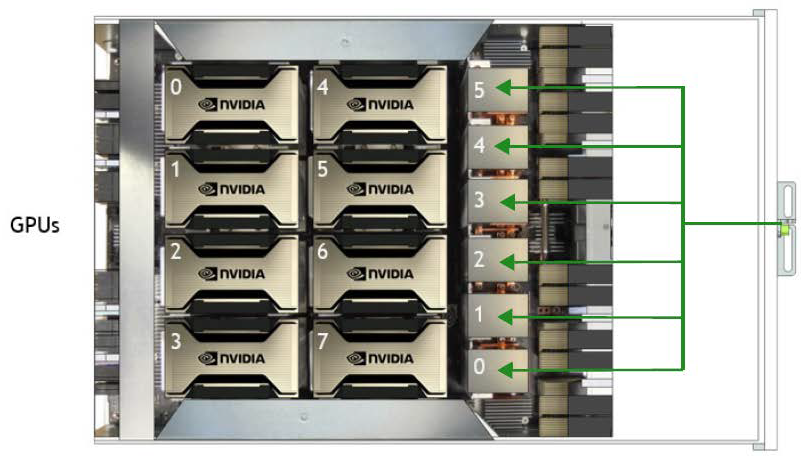}
    \caption{Blower diagram of the DGX A100 GPU server~\cite{dgxguide}.}
    \label{fig:blower}
\end{figure}

Telemetry data is collected via the BMC port. Each server has its own built-in mechanism for managing server operating parameters. Specifically, when the server's operating temperature rises to a threshold of 80 degrees Celsius, the server's automation increases the fan rotation speed to enhance cooling, and when the threshold temperature is exceeded, it reduces power consumption and server performance. The exact algorithms of the server's automation are unknown, but thanks to them, the server maintains operation even with significant performance losses and cooling disruptions. The server's automation does not notify administrators of these operating modes. Entering this mode of operation, among other things, reduces the performance of the HPC cluster when performing parallel calculations. Figure~\ref{fig:blower} shows a blower diagram of the DGX A100 GPU server.

During the operation of the server fleet, the server telemetry data were collected, including:
\begin{itemize}
    \item server lunch times (initial and after repair/maintenance),
    \item server failure times,
    \item GPU utilization,
    \item GPU temperature,
    \item GPU Power consumption,
    \item fan rotation (RPM),
    \item Power Supply Unit (PSU) voltage (3.3V, 5V, and 12V),
    \item CPU utilization,
    \item CPU temperature,
    \item RAM utilization,
    \item and other parameters.
\end{itemize}

The number of analyzed parameter values exceeded 10M, collected over a period of over six months. The data was collected as a multivariate time series, supplemented by information on server downtime periods.

The performed correlation analysis showed that three parameters (GPU utilization, GPU temperature, and GPU Power consumption) provide an effective leading indicator for predicting the increased risk of server failure due to thermal issues. The impact of the other parameters listed above was either significantly less or was sufficient to detect issues using classic threshold-based monitoring.

The physical problem statement was formulated as follows:
\begin{itemize}
    \item an unknown thermal management algorithm (black-box) is implemented in the server using the server automation system.
    \item thermal failures can occur due to internal (fan and automation operation, heat transfer process disruptions, etc.) and environmental factors (power supply, data center ventilation, architectural features, frequent acceleration/deceleration, etc.).
    \item based on utilization, temperature, and power supply consumption data, it is necessary to develop a model that can predict the server's entry into suboptimal operating modes and the increased risk of server failure due to thermal failures, based on historical server operation data.
\end{itemize}

In this formulation, the problem is reduced to constructing a regression model and can be solved using machine learning and quantum machine learning methods, with model training on historical data and subsequent retraining of the models using new data obtained during operation. In this case, to make a management decision on carrying out unscheduled maintenance, the deviation of the actual server operating temperature from the optimal one (predicted based on the results of the model) by a value greater than the threshold was chosen as the metric.

\section{QML Model}
\label{sec:model}

The application of quantum machine learning methods requires the selection of a quantum circuit architecture that optimizes the cost of computing resources for model training and resources during operation, subject to achieving the target performance.

\begin{figure}[htbp]
    \centering
    \includegraphics[width=0.85\linewidth]{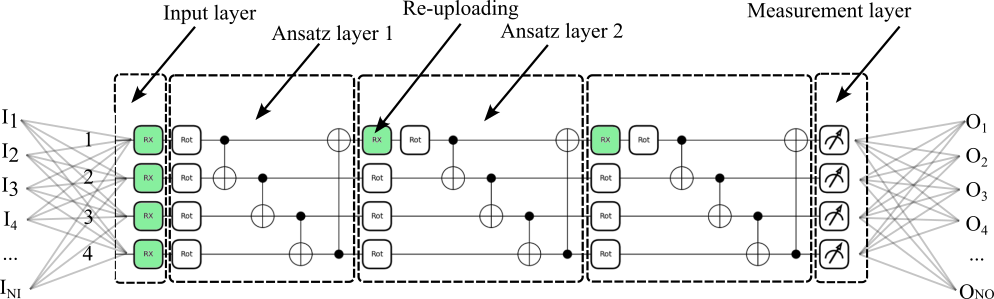}
    \caption{Example of 4 qubits quantum scheme with 3 ansatz layers. It consists of the input layer, few layers of ansatz and the measurement layer. Every ansatz layer consists of rotation gates and entanglement gates. It can include re-uploading gates. Attritionary the scheme includes linear layers on the input and output in order to scaling functions.}
    \label{fig:circuit}
\end{figure}

We based our design on the quantum circuit architecture from~\cite{jerbi2023}, which considers three types of quantum circuits for QML: a simple quantum circuit, a quantum kernel, and a re-uploading circuit. An example of a 4-qubit circuit is shown in Figure~\ref{fig:circuit}. The quantum circuit of their $N$ qubits consists of an input layer, ansatz layers, and a measurement layer. The input layer consists of rotation gates $\Rx(\theta_{i}^{in})$ located on each qubit, i.e., $i = 1, \dots N$. The ansatz plays the role of a learning layer and consists of $\NL$ identical layers. Each layer contains rotation gates at three angles:
\begin{equation}
    Rot = \Rx\left(\theta_{3(i-1) + 3N(j-1) + 1}\right) \Ry\left(\theta_{3(i-1) + 3N(j-1) + 2}\right) \Rz\left(\theta_{3(i-1) + 3N(j-1) + 3}\right),
    \label{eq:rot}
\end{equation}
where $i$ is the qubit number, $j$ is the layer number. The layers contains $N$ control NOT (CNOT) entanglement gates, as shown in the figure. The set and arrangement of gates are identical in each layer. The output of the measurement layer is the average value of each qubit. Since we obtain the data using a quantum computer simulator, there is no need to perform the measurement multiple times and then average it. We use the average value of the gate operator Z obtained from the wave function $m_{i} = \left\langle \psi^{*} \left| Z_{i} \right| \psi \right\rangle$.

Additionally, before each ansatz layer, one additional rotator $\Rx(\theta_{1}^{in})$ can be included with a rotation angle equal to the rotation angle on the zero qubit of the input layer and acting on the first qubit. This scheme is called re-uploading. A scheme without additional rotators in the ansatz will be called a simple quantum circuit.

The gate arguments $\Rx$ in the input layer can only take values between 0 and $2\pi$. In turn, the average measurement values in the output layer can take values between 0 and 1. The arguments and values of the functions we're considering can have a larger range. To overcome this inconsistency, we'll place a single linear neural network layer before the input layer. Similarly, we'll place a linear neural network layer after the output layer. Thus, the angles of the input layer of the quantum circuit are:
\begin{equation}
    \theta_{i}^{in} = \sum_{j=1}^{\NI} w_{i,j}^{in} \cdot I_{j} + b_{i}^{in}.
    \label{eq:input}
\end{equation}
The parameters $w_{i,j}^{in}$ and $b_{i}^{in}$ are responsible for scaling and shifting the input signal $I_{i}$ to the input of the quantum circuit $\theta_{i}^{input}$. On the other hand, the output average value of the Z operators is limited to the interval $[0;1]$, which does not allow us to cover the entire range of function values. Therefore, we also adjust the output values using a linear layer of the neural network:
\begin{equation}
    O_{j} = \sum_{i=1}^{N} w_{j,i}^{out} \cdot m_{i} + b_{j}^{out}.
    \label{eq:output}
\end{equation}
Here, the parameters $w_{j,i}^{out}$ and $b_{j}^{out}$ are also responsible for scaling and offset.

The number of trainable parameters is the sum of the number of rotation angles in the ansatz and the number of linear layer parameters. According to formula~\eqref{eq:rot}, the number of variables $\theta_{i}$ for a circuit with $N$ qubits and $\NL$ layers is $3N\NL$. Additionally, the linear layers of the neural network, according to formulas~\eqref{eq:input} and~\eqref{eq:output}, contribute to the number of parameters $N\NI + N$, where $\NI$ is the number of input parameters, and $N\NO + \NO$, where $\NO$ indicates the number of output parameters. The total number of trainable parameters is:
\begin{equation}
    N_{pq} = \left(3\NL + \NI + \NO\right)N + \NI + \NO = O\left(N\NL\right).
    \label{eq:npq}
\end{equation}
Thus, we obtain a universal approximation system with $I_{j}$ inputs and $O_{j}$ outputs. The $N_{pq}$ number of parameters are used for training. We use common machine learning methods, including gradient descent.

\begin{figure}[htbp]
    \centering
    \includegraphics[width=0.7\linewidth]{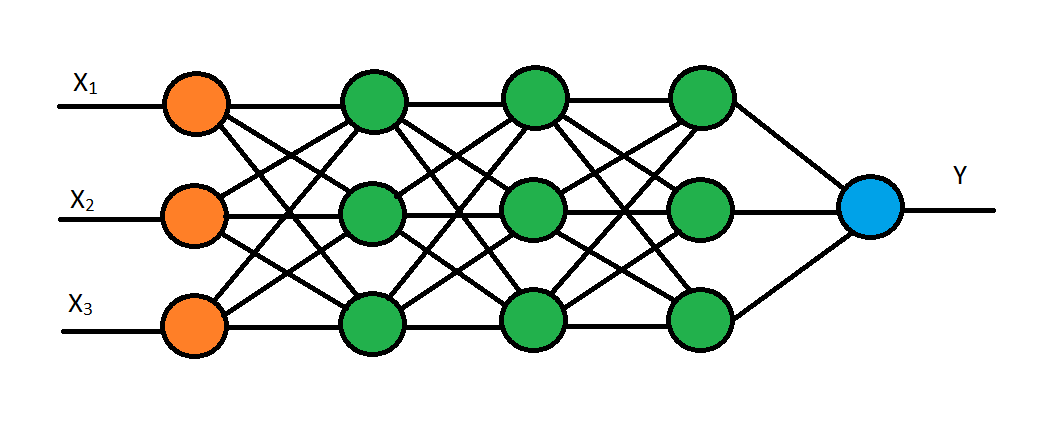}
    \caption{A fully connected neural network with three hidden layers. The input layer is shown in orange, the neurons of the fully connected hidden layers (three neurons per layer) are shown in green, and the output layer is shown in blue.}
    \label{fig:nn}
\end{figure}

The goal of our research is to identify a potential advantages of QML. Based only on analytical approach this task is very complex. Therefore, it is logical to compare a quantum circuit with a similar classical neural network. To compare different approaches within the same conditions, it is useful to consider the number of parameters. Let's consider a classical neural network with $\NL$ layers, with each hidden layer containing N neurons (Figure~\ref{fig:nn}). The input layer contains $\NI$ inputs, and the output layer contains $\NO$ outputs. Then the total number of parameters of a classical neural network is:
\begin{equation}
    N_{pc} = \left(\NL - 2\right)N^{2} + \NI N + N + N\NO + \NO = O\left(\NL N^{2}\right).
    \label{eq:npc}
\end{equation}
In the following steps we will select the number of qubits, neurons and layers so that the values of $N_{pq}$ and $N_{pc}$ are comparable, i.e. the number of trainable parameters is almost the same.

\section{Investigation of Regression on Test Functions}
\label{sec:experiments}

Due the study we performed the computational experiments to confirm the possibility of using quantum machine learning models to solve regression problems and to compare the accuracy of predicting the results of classical fully connected ML models with a sequential number of parameters.

We compare the accuracy of the models for the main classes of functions that describe physical processes:
\begin{itemize}
    \item harmonic (periodic) function with damping of the form:
    \begin{equation}
        y = \sin\left(\left(\frac{x^{2}}{10}\right)\right) \cdot \frac{5}{1+x},
        \label{eq:harmonic}
    \end{equation}
    \item polynomial function of the form
    \begin{equation}
        y = \frac{x^{7} - 14x^{5} + 49x^{3} - 36x}{100},
        \label{eq:poly}
    \end{equation}
    \item linear-theta function (discontinuous non-differentiable piecewise-defined function) ($x < 5$, $y = 0.1*x$; $x \geq 5$, $y = x$)
    \begin{equation}
        y = 0.1x\Theta\left(5-x\right) + x\Theta\left(x-5\right).
        \label{eq:theta}
    \end{equation}
\end{itemize}

For our experiments we defined functions on the interval $[0;10]$. The input and output (Figure~\ref{fig:circuit}) were $\NI = 1$, $\NO = 1$.

The performance of the models was compared pairwise for models with 2, 3, 4, 5, 6, 8 and 10 neurons/qubits per layer and for 2, 3, 4, 5, 6, 8 and 10 ansatz layers/hidden layers of the neural network, each with an equal or comparable number of parameters. Data on the number of parameters of the quantum circuit and neural network depending on the number of neurons/qubits and the number of layers are given by formulas~\eqref{eq:npq} and~\eqref{eq:npc}.

We used MAD (Mean Absolute Difference) metrics for performance comparation.

In the next stage of the study we compared a simple quantum circuit, a quantum circuit with re-uploading, and a fully connected neural network with a hyperbolic tanh activation function. We didn't injected the quantum noise (Figure~\ref{fig:circuit}) till the simulation (fidelity = 1).

\subsection{Investigation of Regression on Periodic Functions}
\label{sec:periodic}

Simulation for periodic functions similar to the function~\eqref{eq:harmonic} showed that the quantum circuit shows better performance in comparison to a fully connected neural network.

\begin{figure}[htbp]
    \centering
    \includegraphics[width=0.7\linewidth]{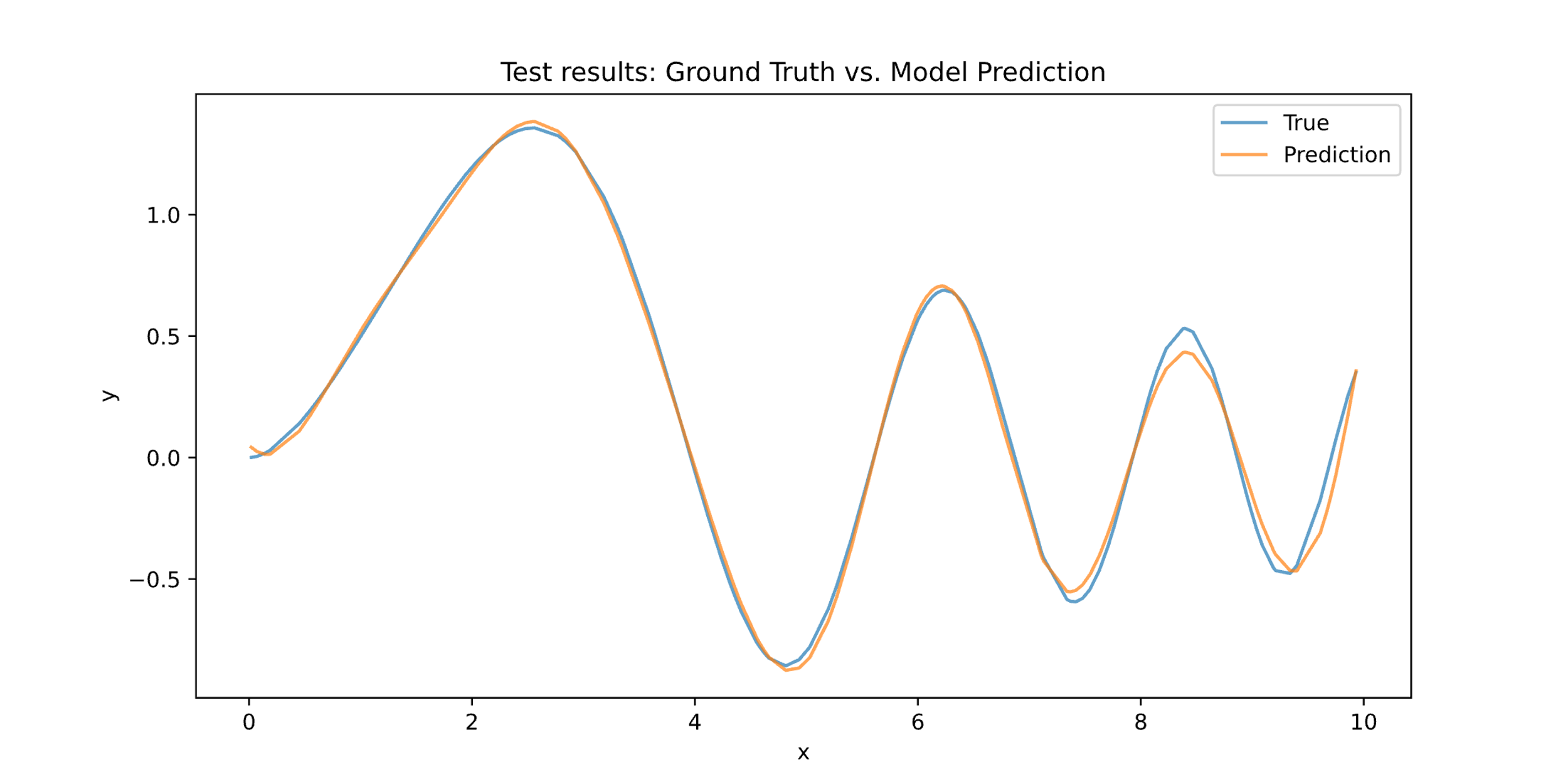}
    \caption{A harmonic function predicted by the quantum circuit with 4 qubits and 4 layers. It's used quantum circuit with re-uploading.}
    \label{fig:harm-qc}
\end{figure}

\begin{figure}[htbp]
    \centering
    \includegraphics[width=0.7\linewidth]{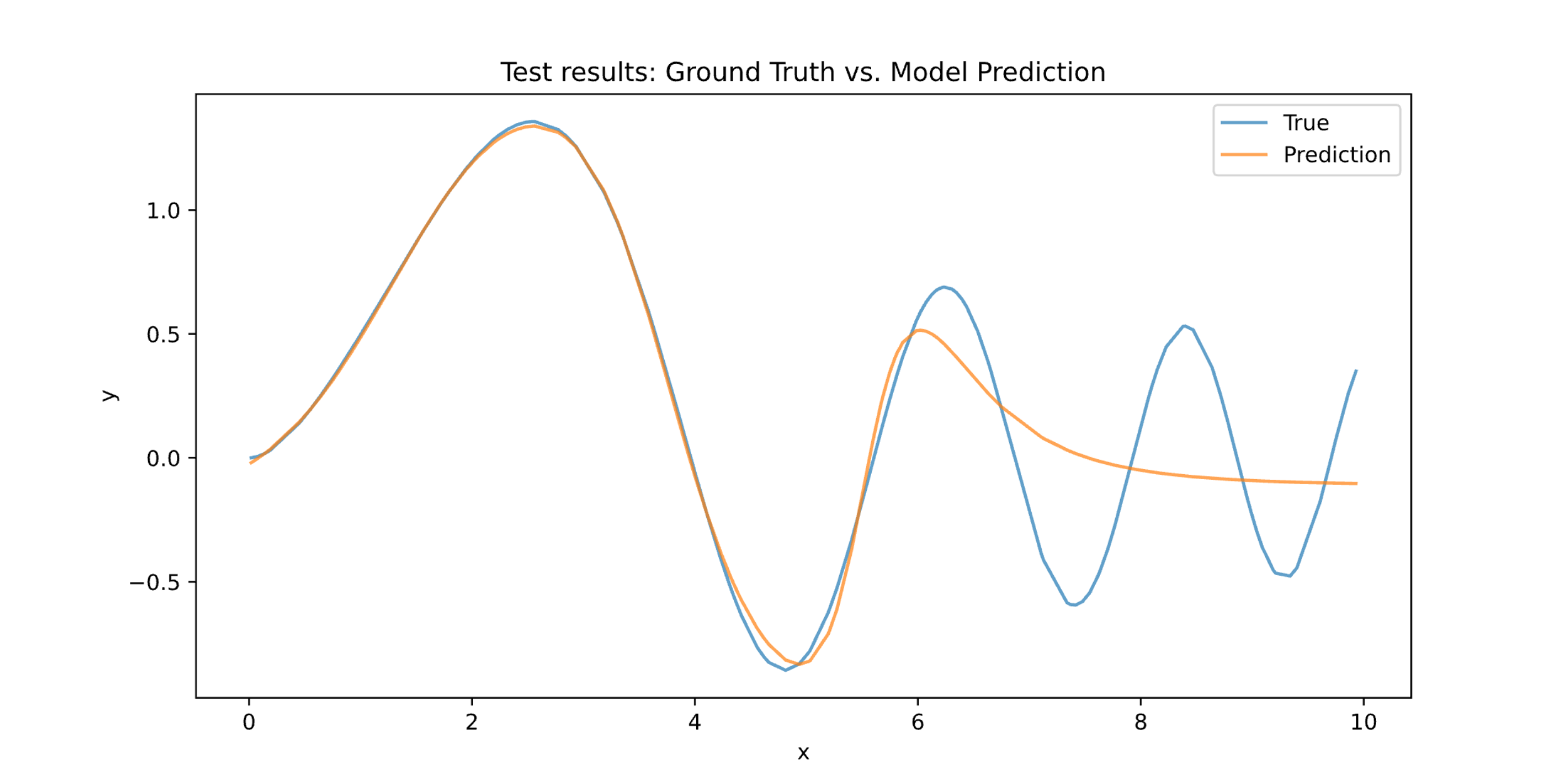}
    \caption{A harmonic function predicted by neural network with 4 neurons per layer and 4 hidden layers.}
    \label{fig:harm-nn}
\end{figure}

\begin{figure}[htbp]
    \centering
    \includegraphics[width=0.7\linewidth]{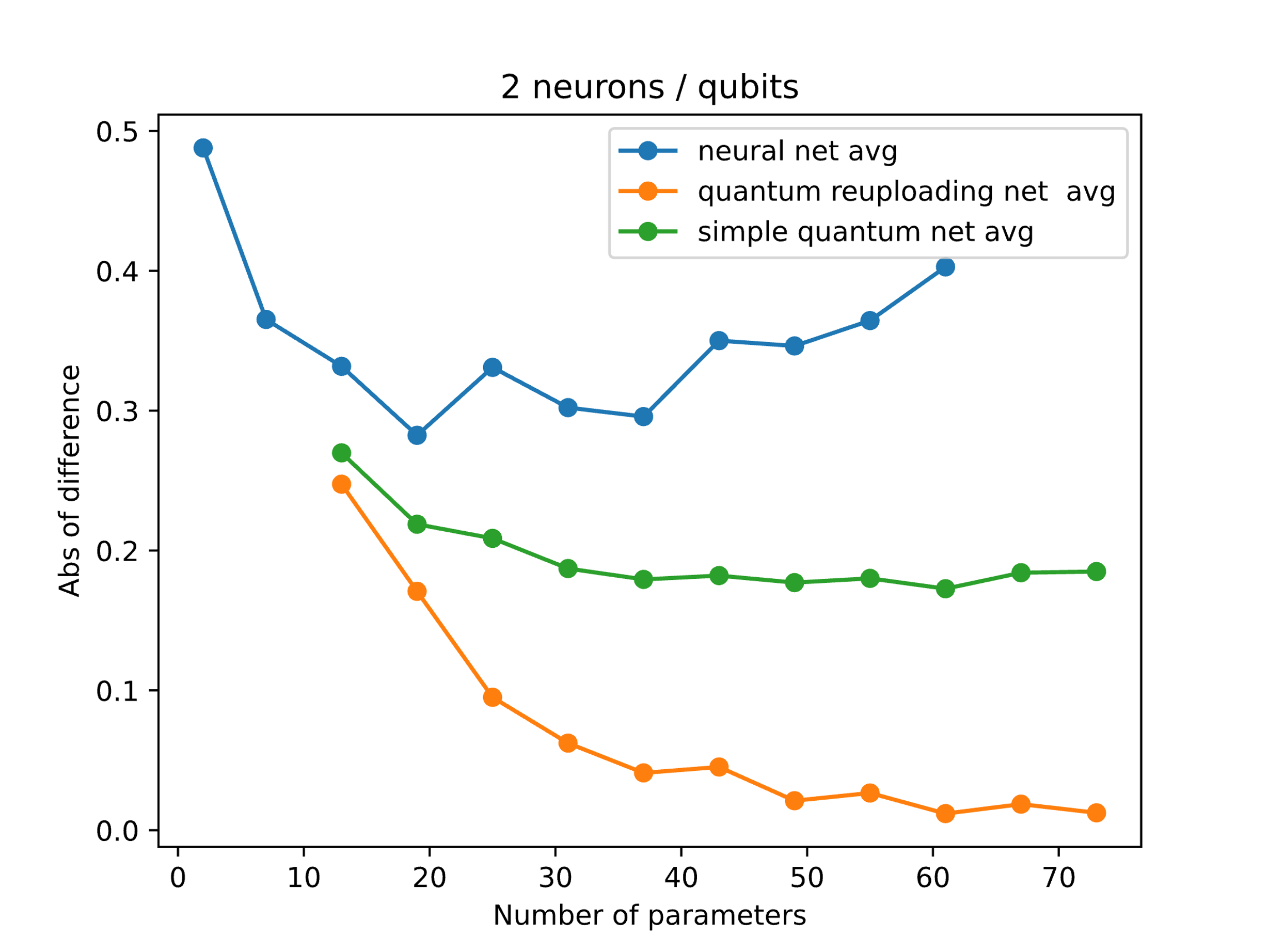}
    \caption{Comparation 2 qubit circuit and 2 neuron per layer for harmonic function.}
    \label{fig:harm-cmp2}
\end{figure}

\begin{figure}[htbp]
    \centering
    \includegraphics[width=0.7\linewidth]{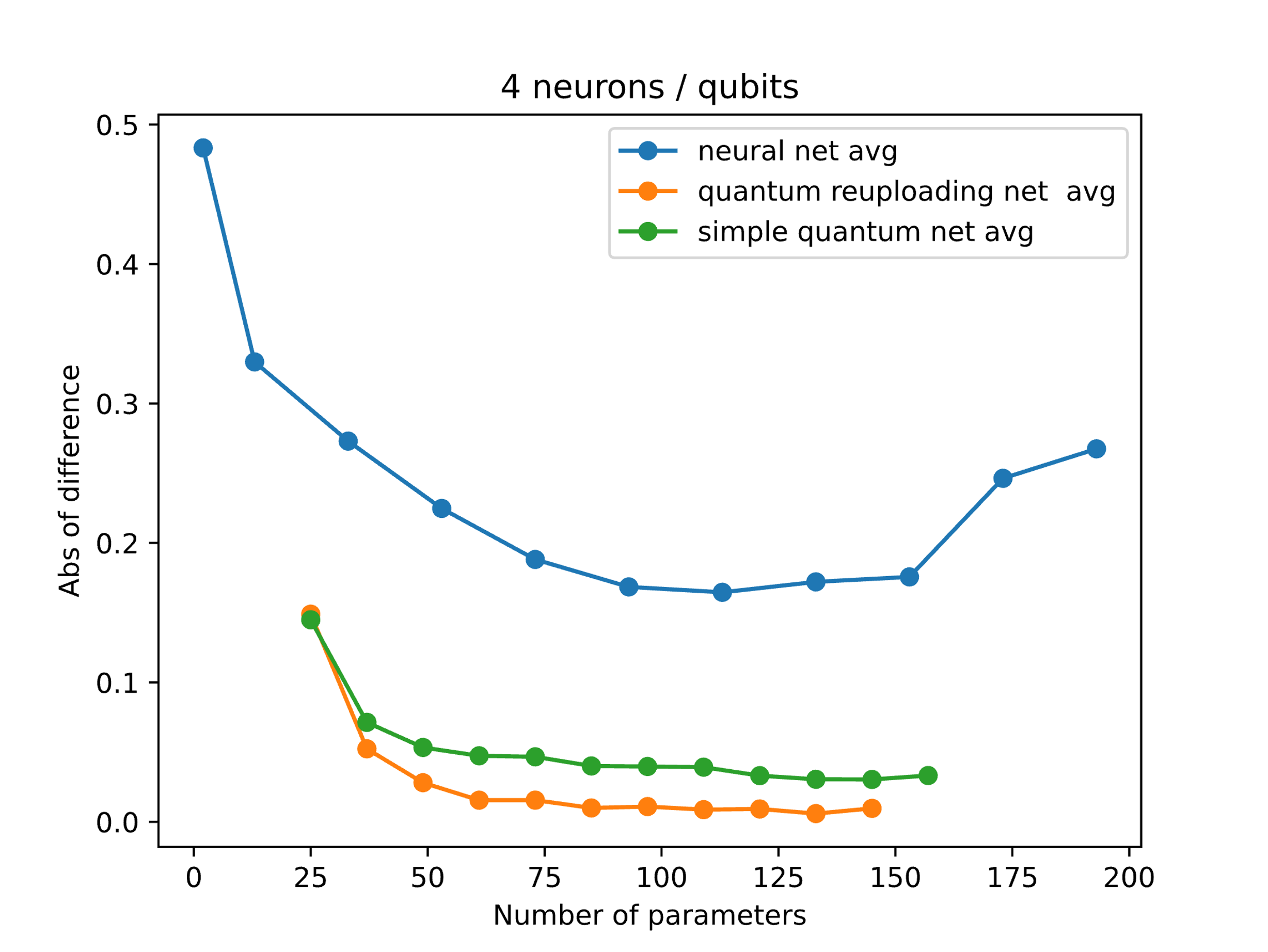}
    \caption{Comparation of performance for 4 qubits circuit and 4 neurons per hidden layer for harmonic function.}
    \label{fig:harm-cmp4}
\end{figure}

\begin{figure}[htbp]
    \centering
    \includegraphics[width=0.7\linewidth]{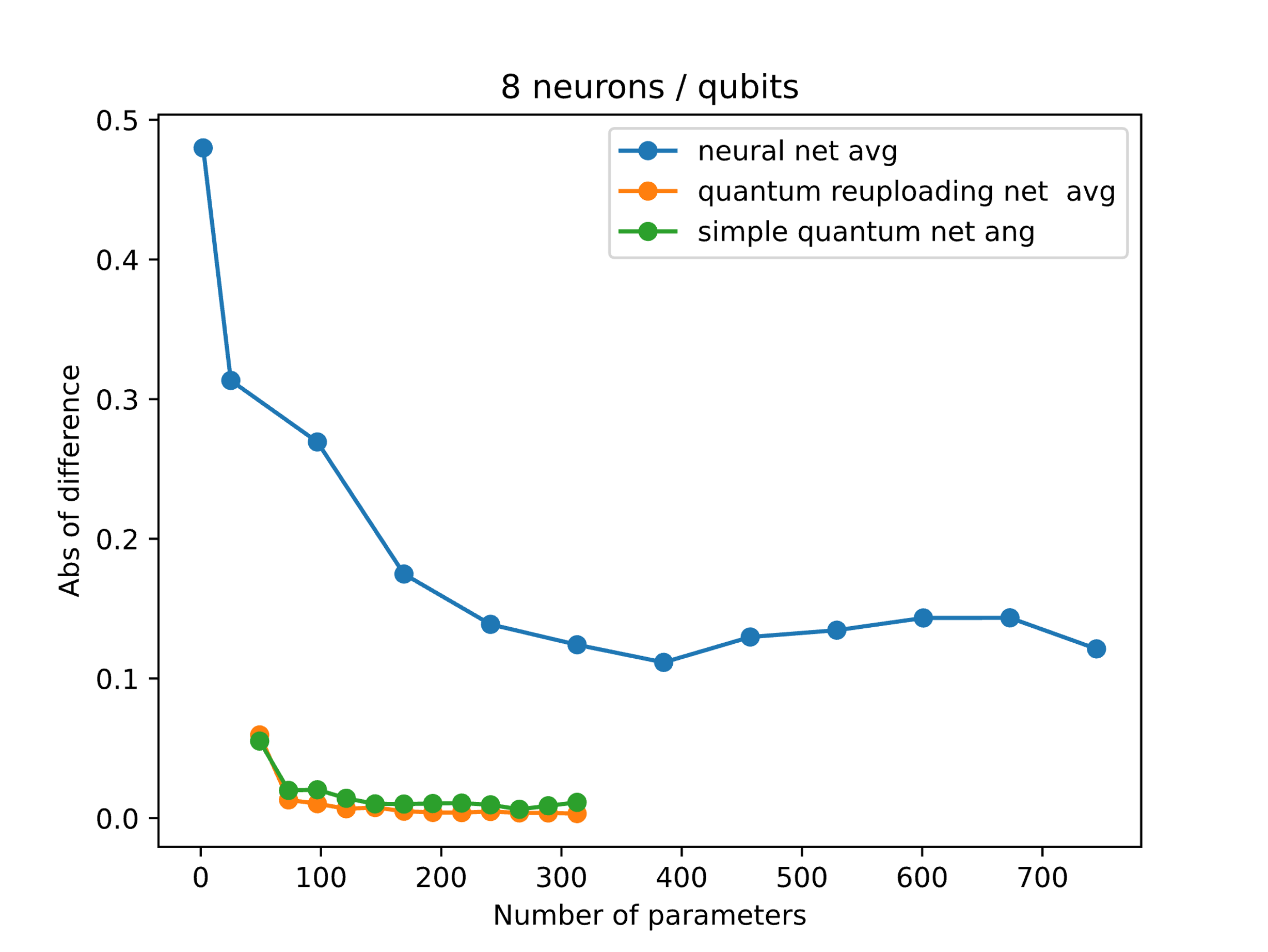}
    \caption{Comparation of performance 8 qubit circuit and 8 neurons per layer for harmonic function.}
    \label{fig:harm-cmp8}
\end{figure}

The results of approximation of the harmonic function by the quantum circuit and neural network are shown in Figure~\ref{fig:harm-qc} and Figure~\ref{fig:harm-nn}. The results of comparison of the prediction accuracy are shown in Figure~\ref{fig:harm-cmp2}~--~Figure~\ref{fig:harm-cmp8}.

When applied to modeling periodic functions, computational experiments have shown an accuracy advantage for a quantum circuit with a re-uploading architecture compared to a simple quantum circuit, as well as an improvement in performance with an increasing number of ansatz layers and qubits.

\subsection{Investigation of Regression on Polynomial Functions}
\label{sec:polynomial}

\begin{figure}[htbp]
    \centering
    \includegraphics[width=0.7\linewidth]{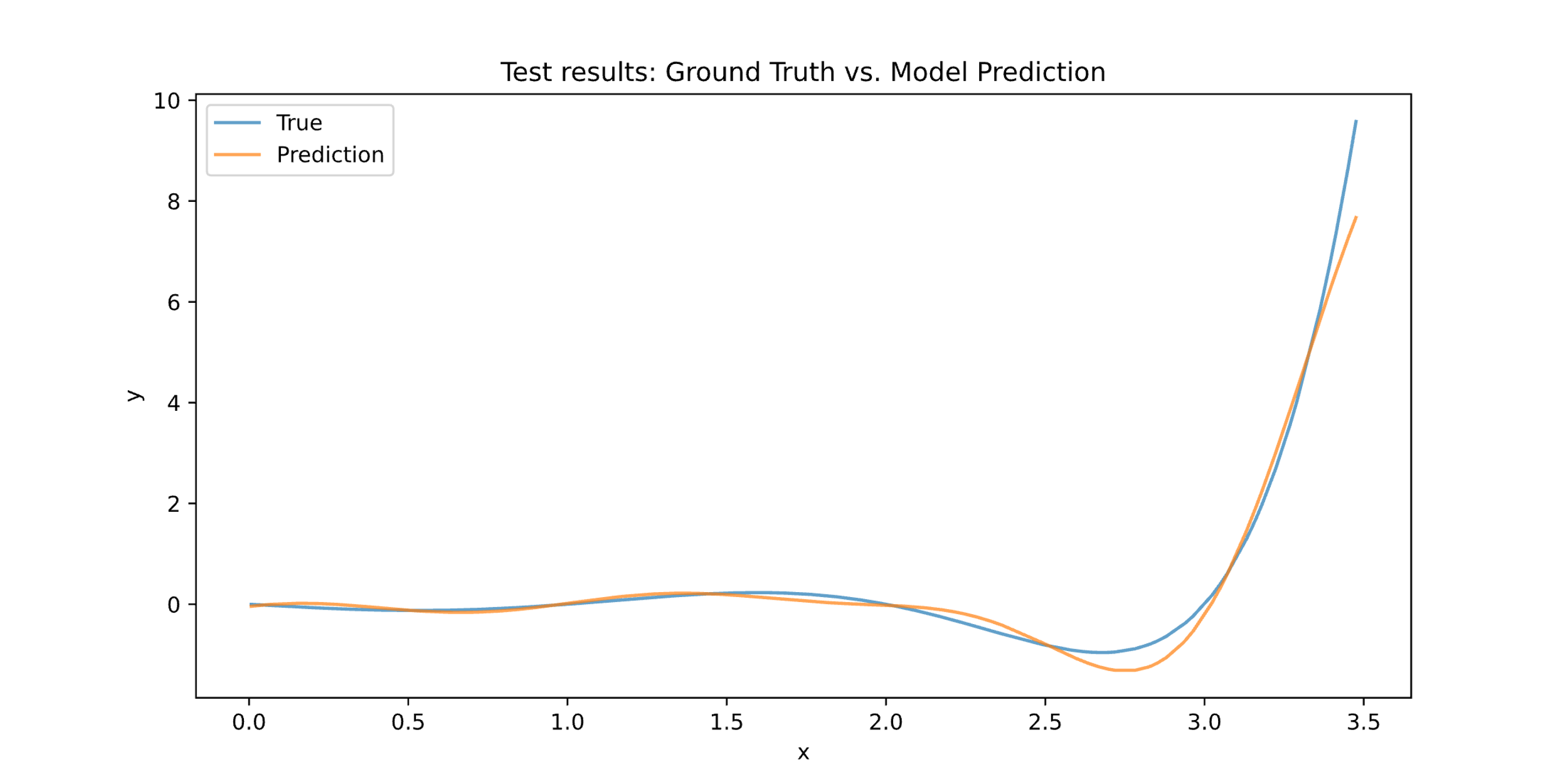}
    \caption{A polynomic function predicted by the quantum circuit with 4 qubits and 6 layers with re-uploading.}
    \label{fig:poly-qc}
\end{figure}

\begin{figure}[htbp]
    \centering
    \includegraphics[width=0.7\linewidth]{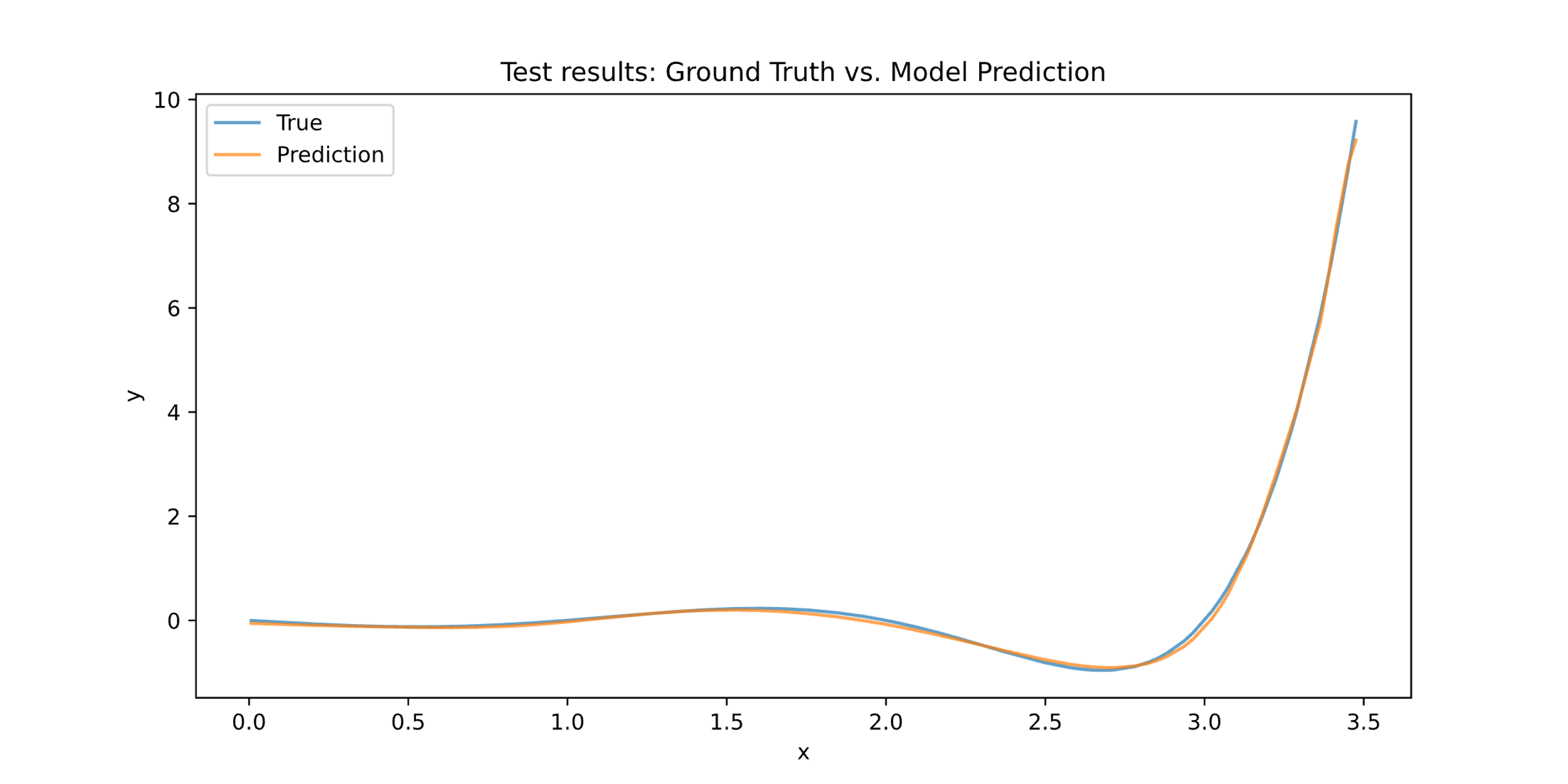}
    \caption{A polynomic function predicted by the fully connected neural network with 4 neurons and 6 layers.}
    \label{fig:poly-nn}
\end{figure}

\begin{figure}[htbp]
    \centering
    \includegraphics[width=0.7\linewidth]{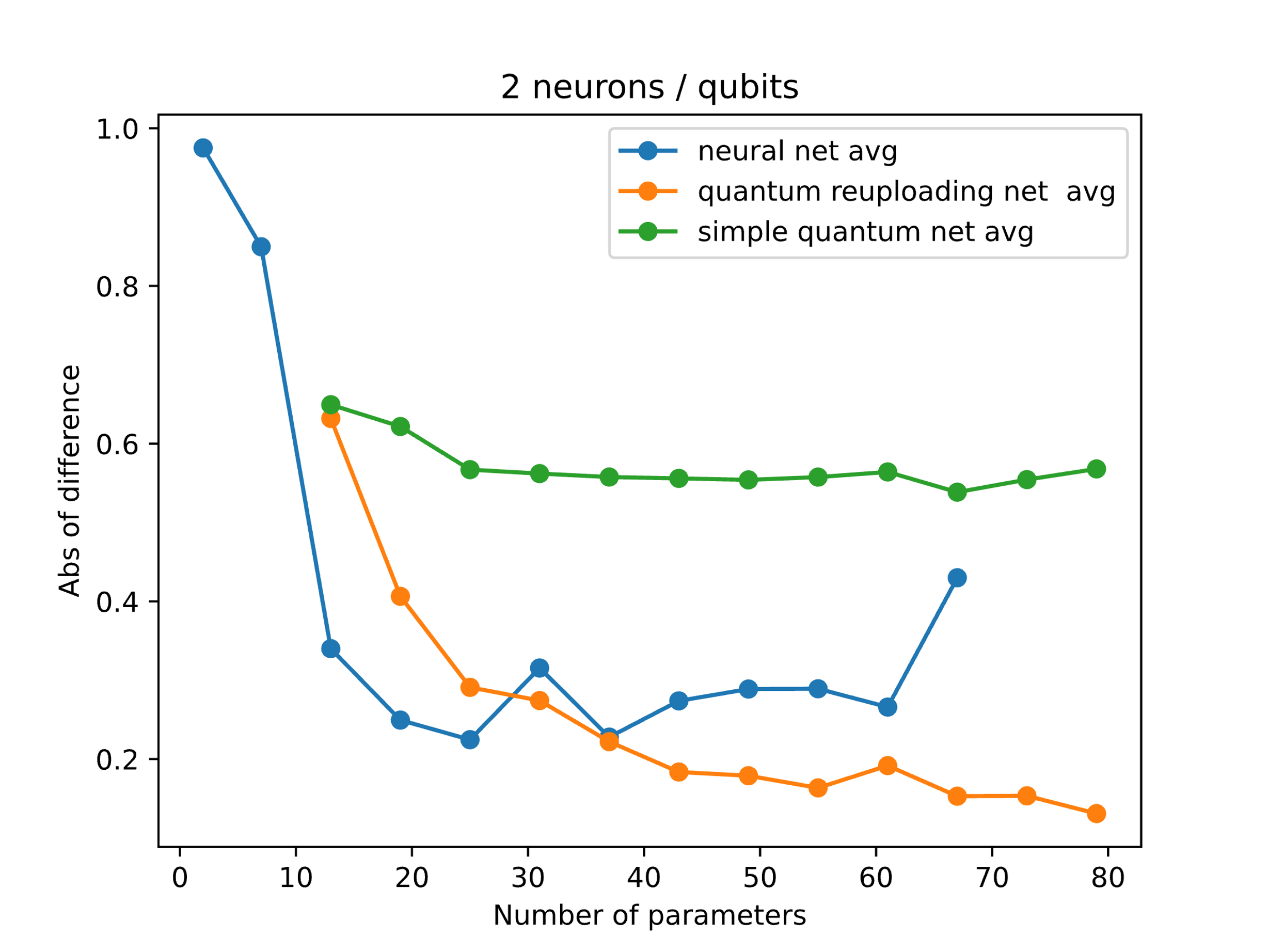}
    \caption{Comparation 2 qubit circuit and 2 neurons per layer for polynomic function.}
    \label{fig:poly-cmp2}
\end{figure}

\begin{figure}[htbp]
    \centering
    \includegraphics[width=0.7\linewidth]{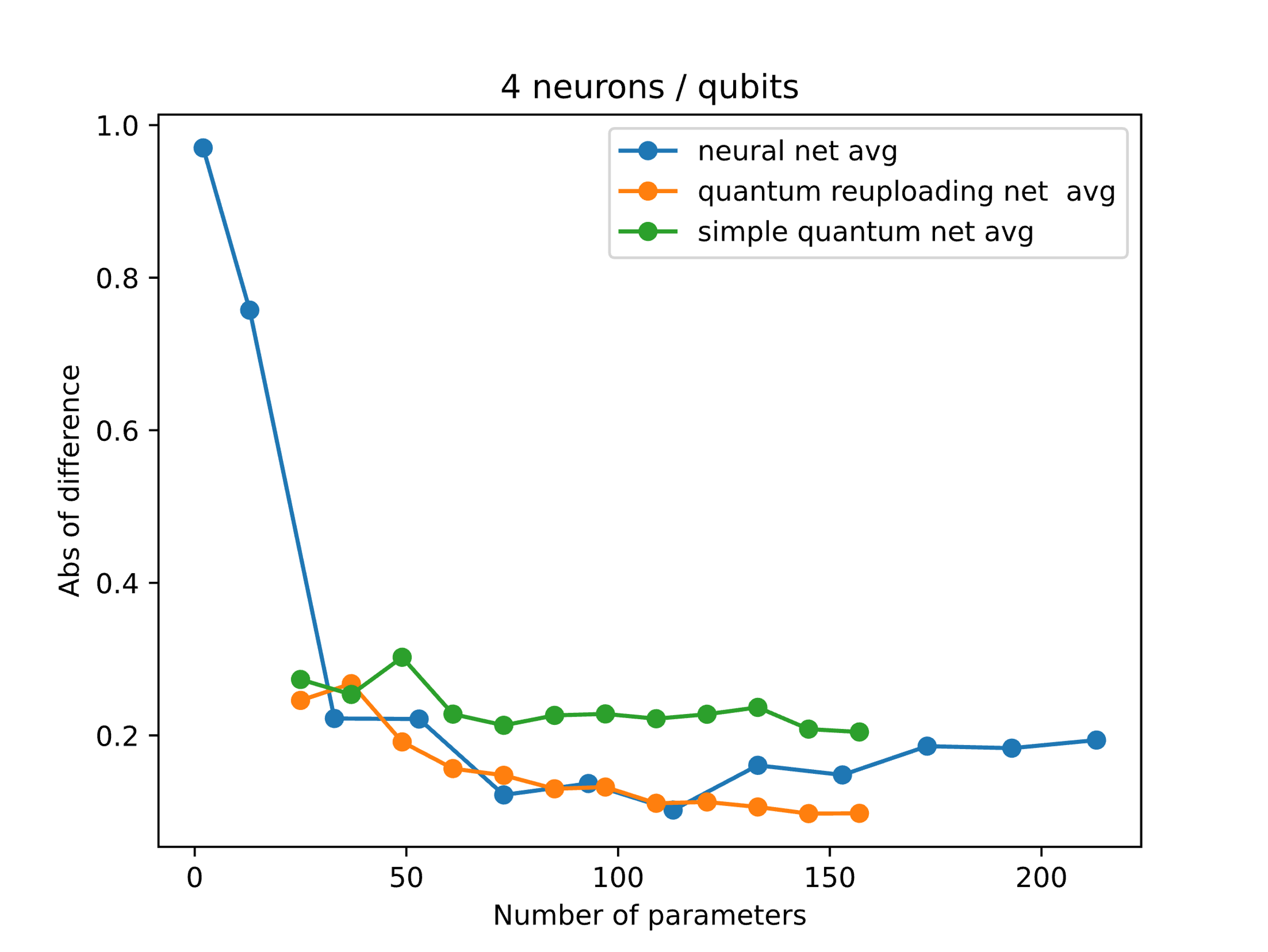}
    \caption{Comparation 4 qubit circuit and 4 neurons per layer for polynomic function.}
    \label{fig:poly-cmp4}
\end{figure}

\begin{figure}[htbp]
    \centering
    \includegraphics[width=0.7\linewidth]{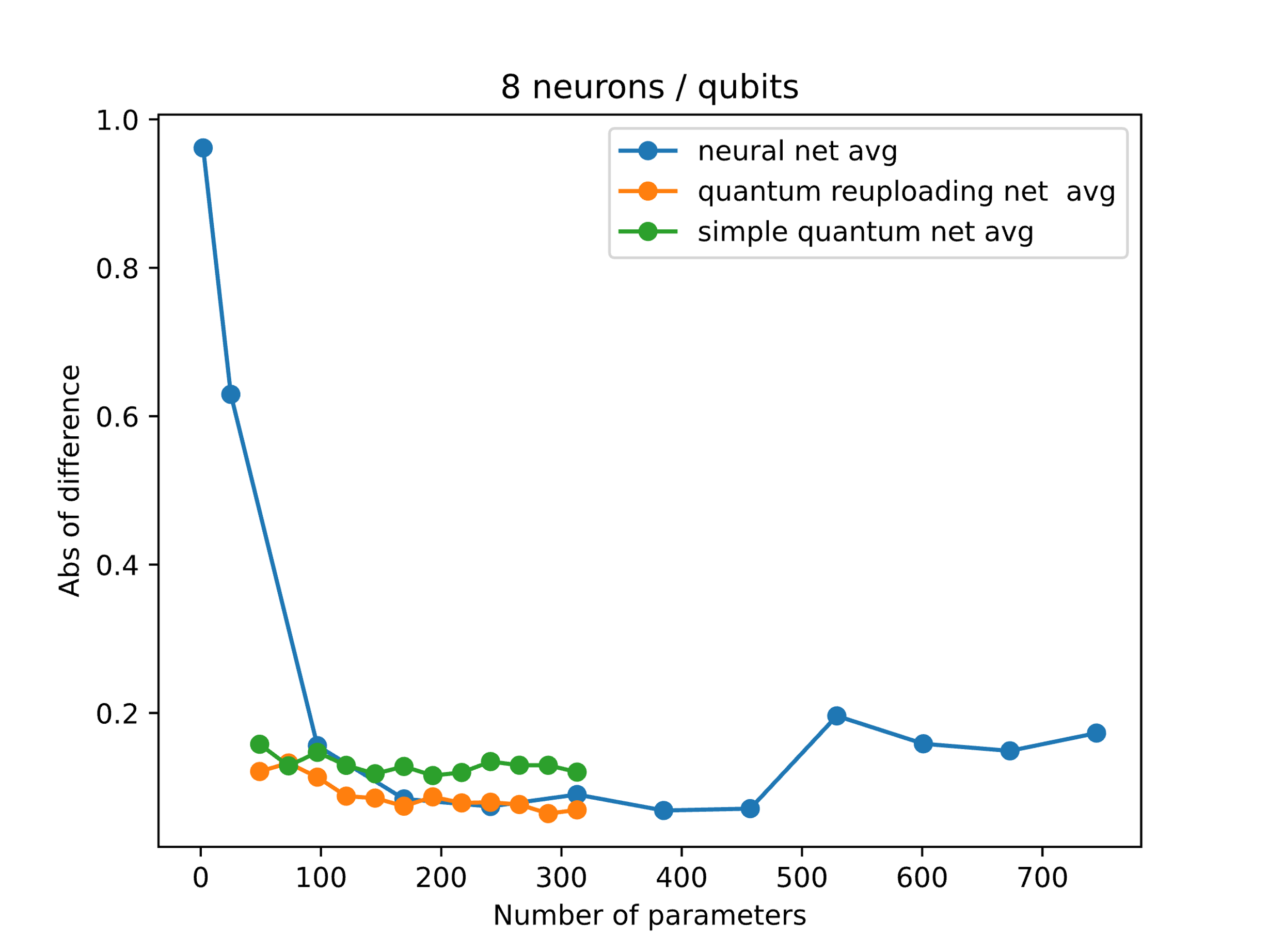}
    \caption{Comparation 8 qubit circuit and 8 neurons per layer for polynomic function.}
    \label{fig:poly-cmp8}
\end{figure}

Simulations for a polynomial function of the form~\eqref{eq:poly} showed comparable accuracy when performed by a quantum circuit and a neural network. The examples of approximations for polynomial function are shown in Figure~\ref{fig:poly-qc}~--~Figure~\ref{fig:poly-nn}. The approximation accuracy during simulations are shown in Figure~\ref{fig:poly-cmp2}~--~Figure~\ref{fig:poly-cmp8}.

As the number of parameters increases when modeling a polynomial function, the accuracy of the neural network and the quantum circuit with re-uploading increases and converges. However, the neural network loses its advantage as the number of network parameters increases. The re-uploading architecture also demonstrated a significant advantage over the simple quantum circuit.

\subsection{Investigation of Regression on Linear-Theta Functions}
\label{sec:theta}

\begin{figure}[htbp]
    \centering
    \includegraphics[width=0.7\linewidth]{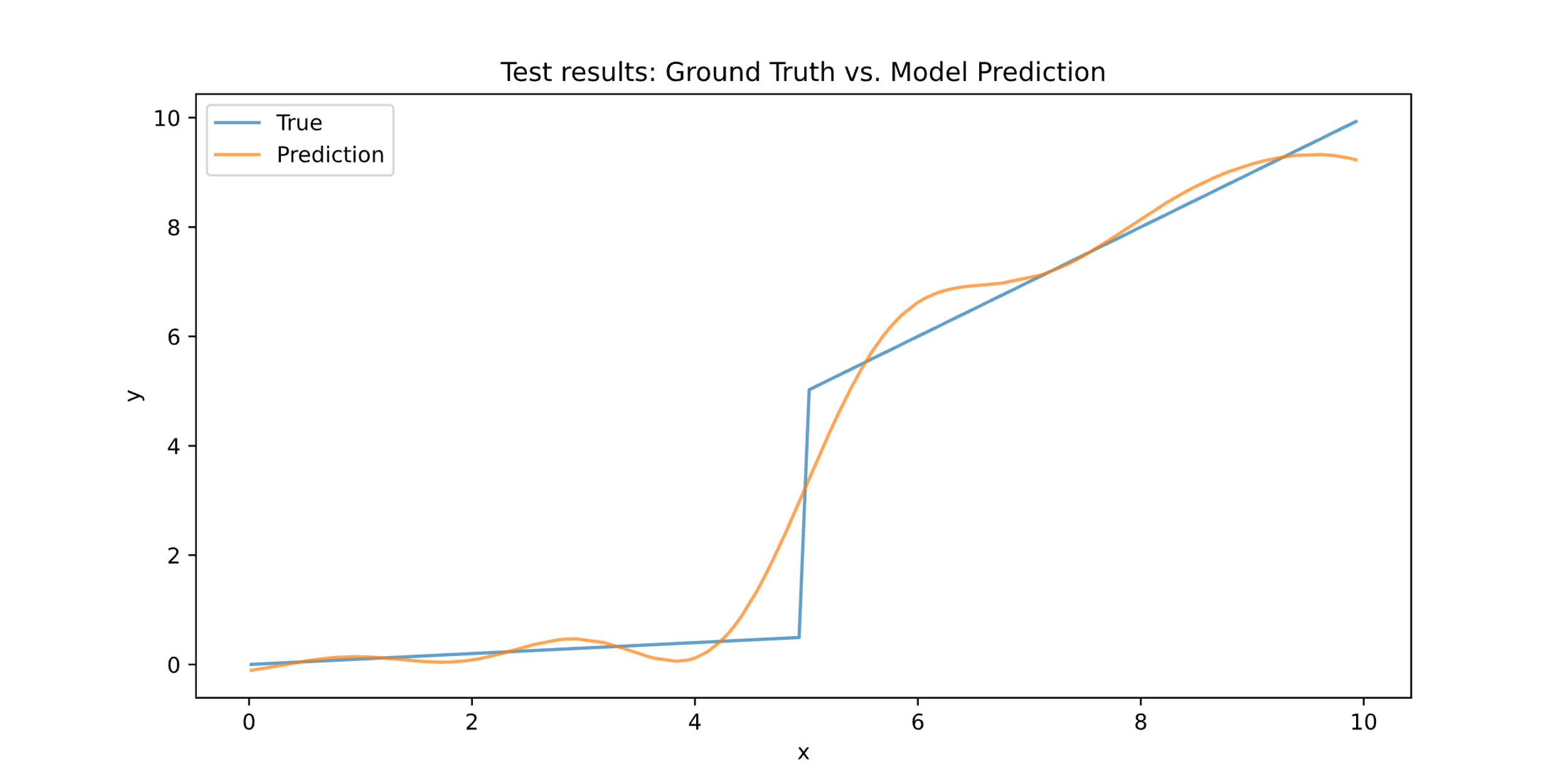}
    \caption{A linear-theta function predicted by the quantum circuit with 4 qubits and 5 layers. Re-uploading scheme is used.}
    \label{fig:theta-qc}
\end{figure}

\begin{figure}[htbp]
    \centering
    \includegraphics[width=0.7\linewidth]{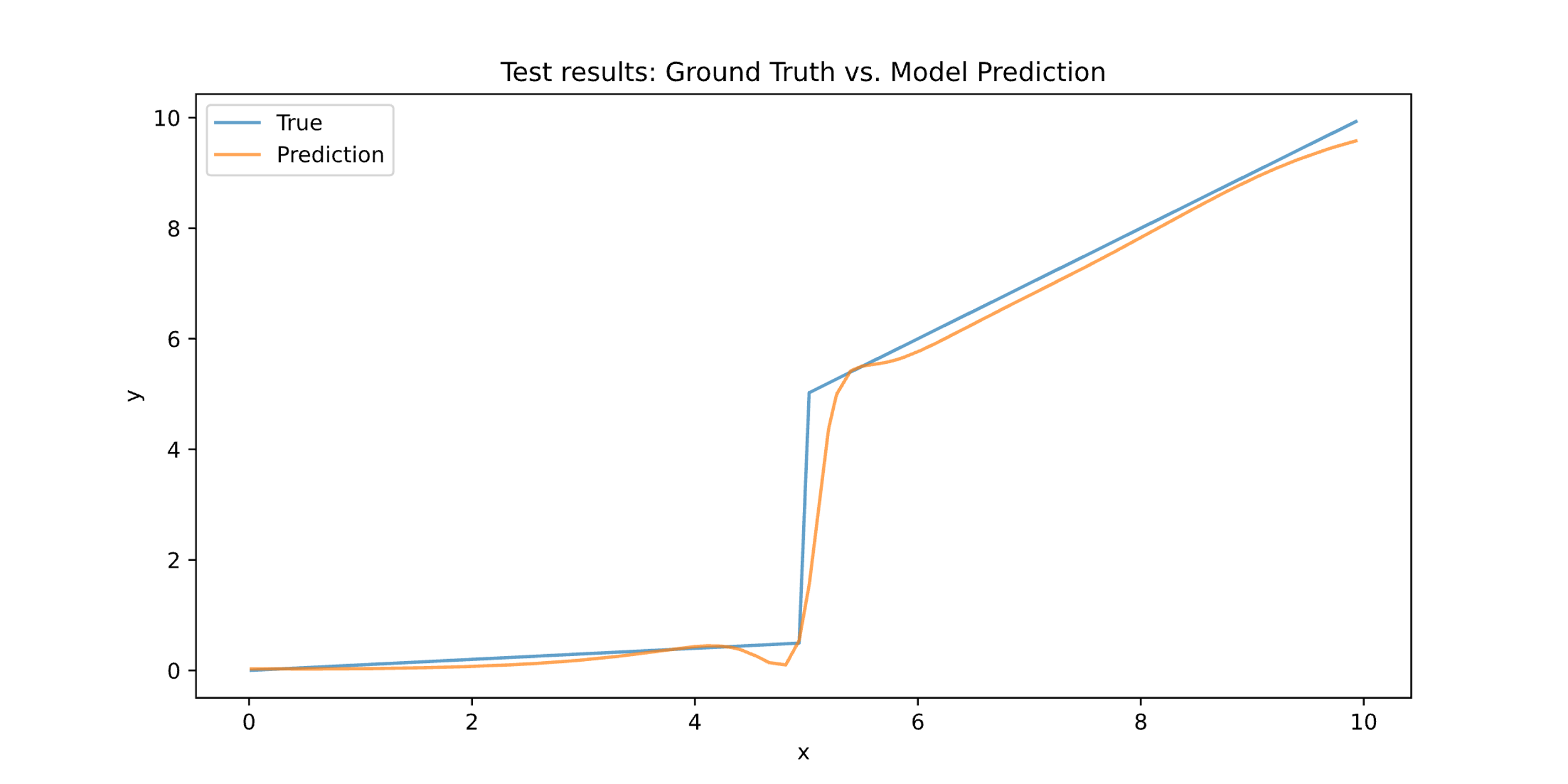}
    \caption{A linear-theta function predicted by the neural network with 4 neurons and 5 hidden layers.}
    \label{fig:theta-nn}
\end{figure}

\begin{figure}[htbp]
    \centering
    \includegraphics[width=0.7\linewidth]{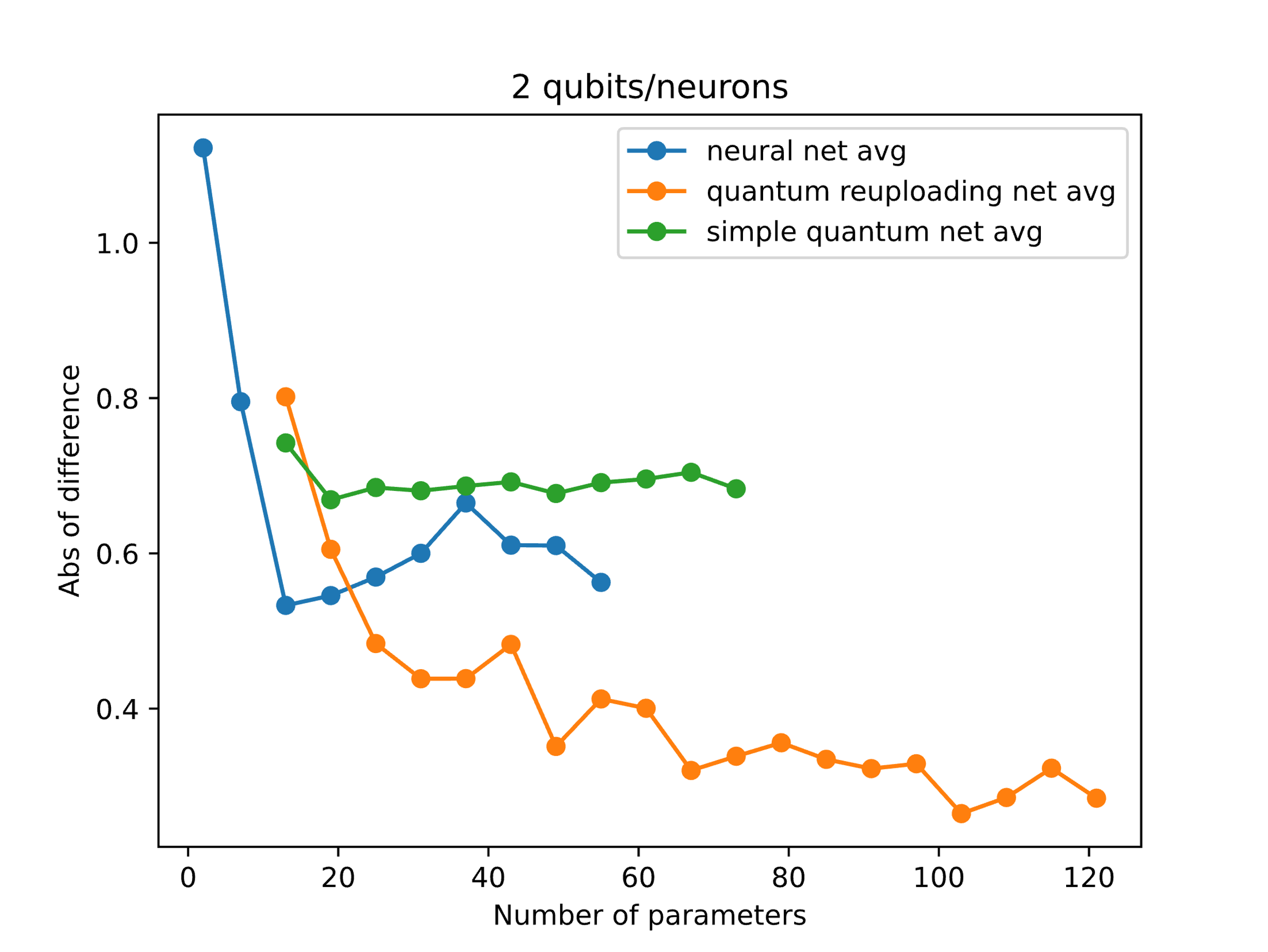}
    \caption{Comparation 2 qubits circuit and 2 neurons per layer for a linear-theta function.}
    \label{fig:theta-cmp2}
\end{figure}

\begin{figure}[htbp]
    \centering
    \includegraphics[width=0.7\linewidth]{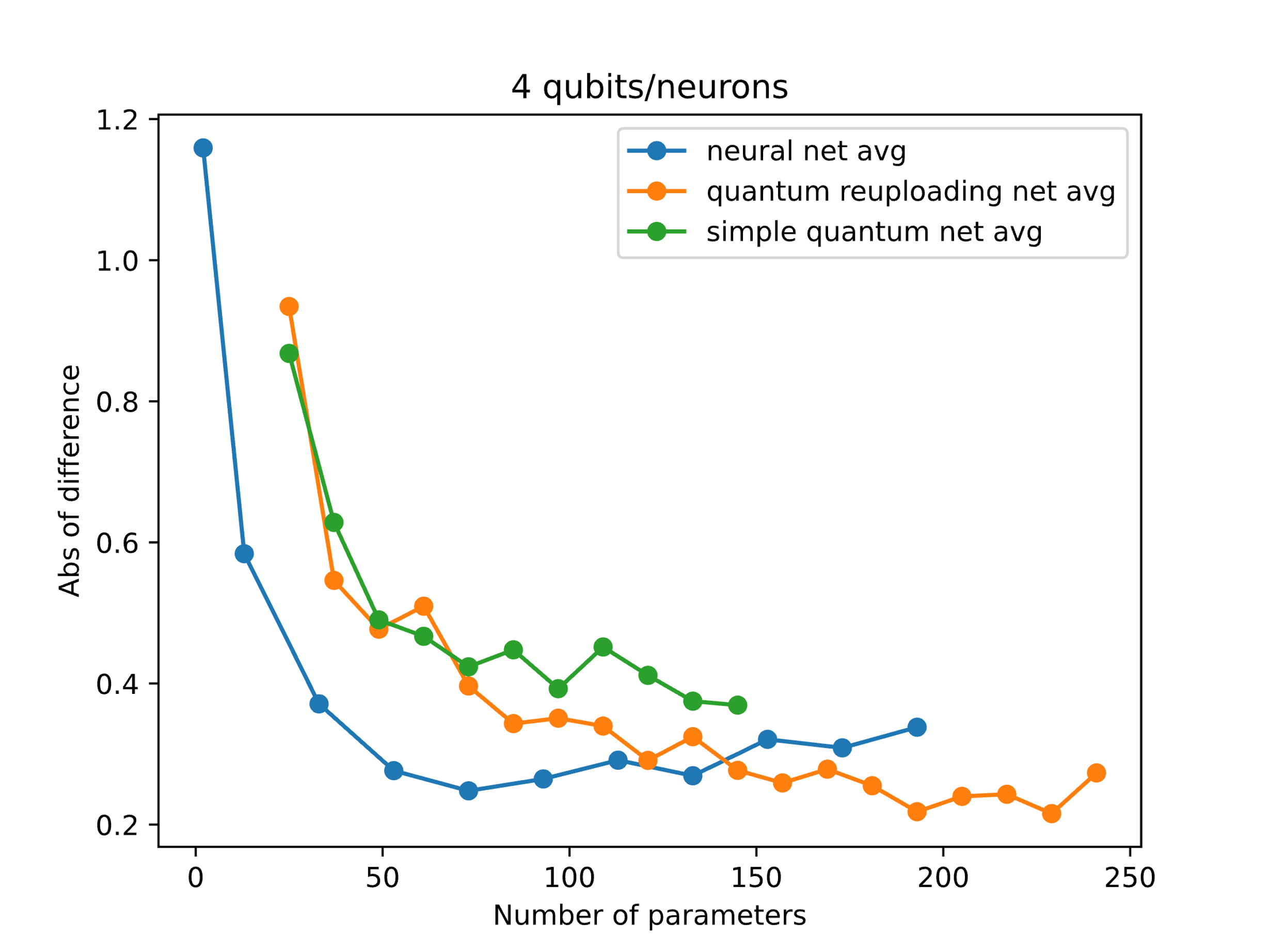}
    \caption{Comparation 4 qubits circuit and 4 neurons per layer for a linear-theta function.}
    \label{fig:theta-cmp4}
\end{figure}

\begin{figure}[htbp]
    \centering
    \includegraphics[width=0.7\linewidth]{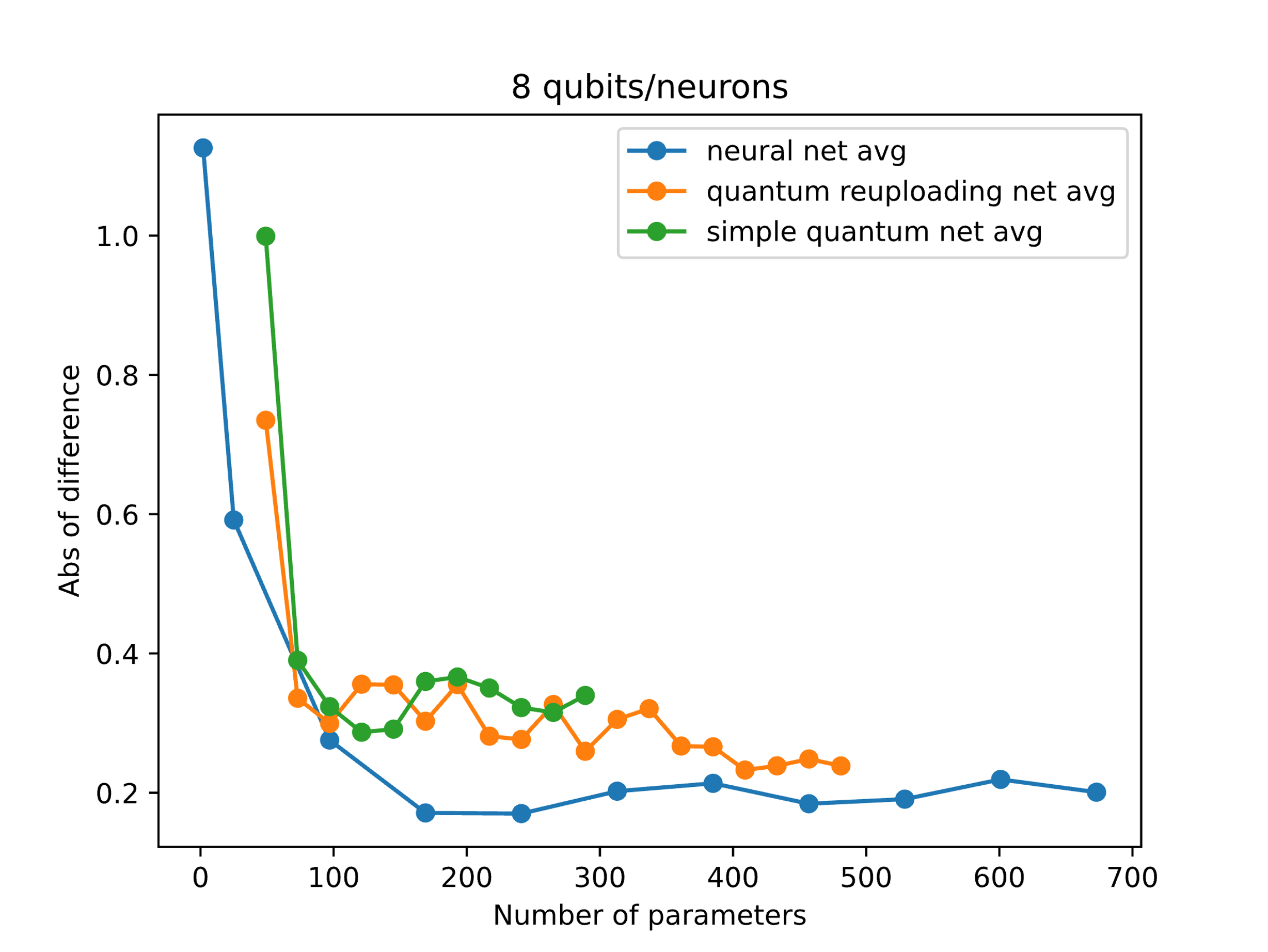}
    \caption{Comparation 8 qubit circuit and 8 neurons per layer for a linear-theta function.}
    \label{fig:theta-cmp8}
\end{figure}

Simulation of the linear-theta function~\eqref{eq:theta} also demonstrated comparable accuracy due to performing of the quantum circuit and the neural network. Examples of piecewise linear function approximations are shown in Figure~\ref{fig:theta-qc} and Figure~\ref{fig:theta-nn}. Comparisons of approximation accuracy are shown in Figure~\ref{fig:theta-cmp2}~--~Figure~\ref{fig:theta-cmp8}.

As with a polynomial function, as the number of parameters increases when modeling a step function, the accuracy of the neural network and quantum circuit with re-uploading increases and converges until saturation is reached with increasing number of network parameters. The re-uploading architecture for this class of functions also demonstrated a significant advantage over a simple quantum circuit.

\section{Results}
\label{sec:results}

To predict server GPU temperatures, we used the power consumption telemetry data for each of the 8 GPUs of each server in the cluster $P_{j}$ and utilization data $U_{j}$ as input. The size of the QML input vector is $\NI = 16$. This one predicts 8 temperatures, separately for each GPU, hence $\NO = 8$ (Figure~\ref{fig:circuit}). The telemetry dataset contains 270,000 parameter measurements. For training, we used a random sample of 10\% of the original dataset. During model training process we split the dataset into train (70\%), validation (15\%), and test data (15\%). For validation process we used the MSE (Mean Squared Error) loss function. The final evaluation of training performance on the test dataset was estimated as the MAD of all predicted and real temperatures:
\begin{equation}
    MAD = \frac{1}{8N_{test}} \sum_{i=1}^{N_{test}} \sum_{j=1}^{8} \left| T^{i,j}_{true} - T^{i,j}_{pred} \right|,
    \label{eq:mad}
\end{equation}
where $N_{test}$ -- the number of records in the test dataset.

\begin{figure}[htbp]
    \centering
    \includegraphics[width=0.7\linewidth]{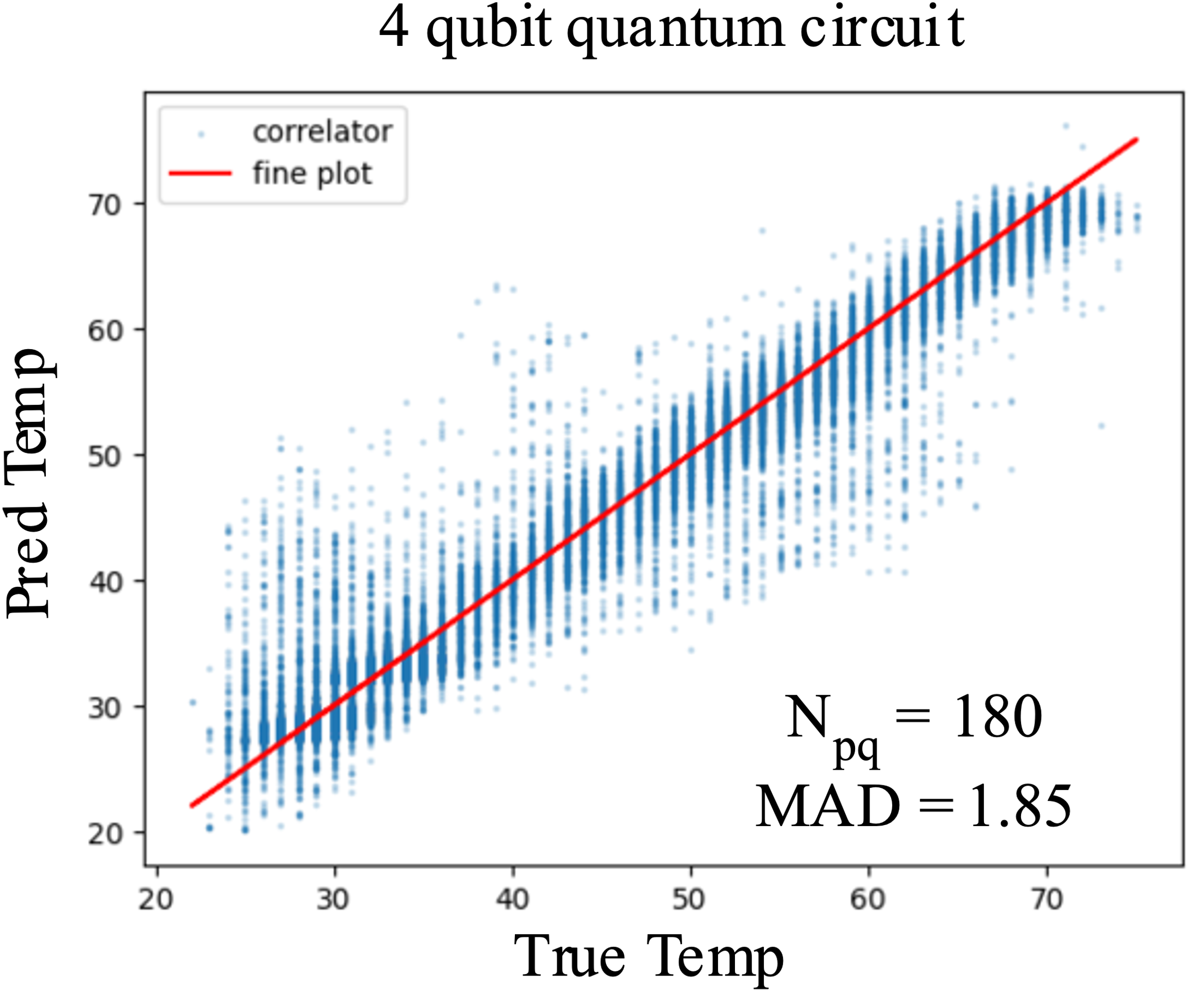}
    \caption{Correlation -- dependence of predicted temperature on true temperature. Fine plot depicts true temperature. Quantum circuit consists of 4 qubits and 6 layers of ansatz.}
    \label{fig:corr-qc4}
\end{figure}

\begin{figure}[htbp]
    \centering
    \includegraphics[width=0.7\linewidth]{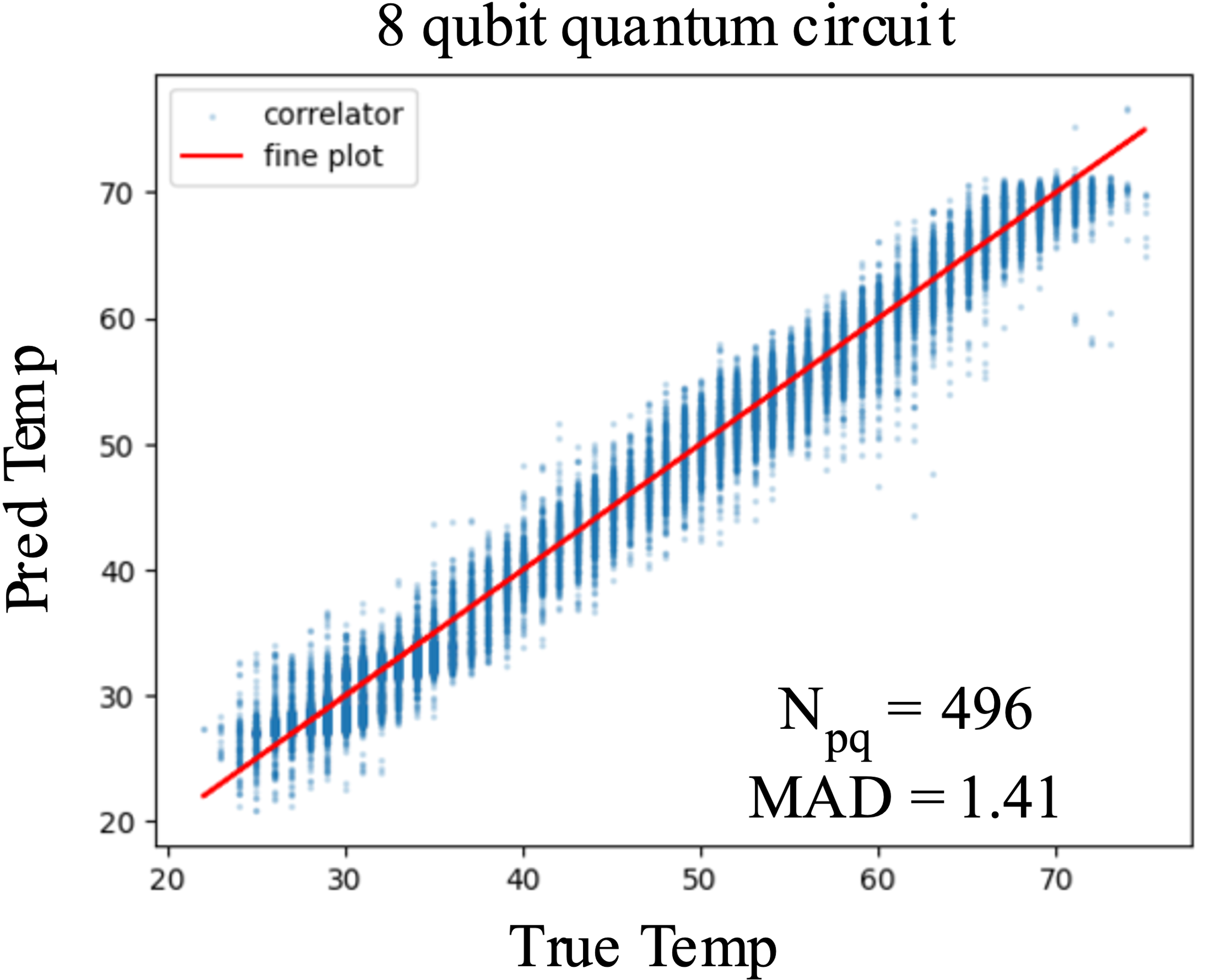}
    \caption{Correlation -- dependence of predicted temperature on true temperature. Fine plot depicts true temperature. Quantum circuit consists of 8 qubits and 12 layers of ansatz.}
    \label{fig:corr-qc8}
\end{figure}

The temperature prediction results are shown in Figure~\ref{fig:corr-qc4}~--~Figure~\ref{fig:corr-qc8}. The blue dots in the figures show the correlations between the predicted temperature and the real temperature taken from the dataset. The red line indicates the real temperature. If the dot is closer to the red line, then the correspondent temperature prediction is better. Figure~\ref{fig:corr-qc4} shows the prediction results for a 4-qubit circuit with 4 ansatz layers and 180 parameters. The average deviation is MAD = 1.85~\degC. At the same time we observed a wide spread of predicted temperatures. The difference between max and min predicted temperature is 33\degC. An 8-qubit quantum circuit with 12 layers performs significantly better in temperature prediction (Figure~\ref{fig:corr-qc8}). This configuration has 496 parameters. The average deviation is MAD = 1.41~\degC. The maximum deviation is 15\degC.

\begin{figure}[htbp]
    \centering
    \includegraphics[width=0.7\linewidth]{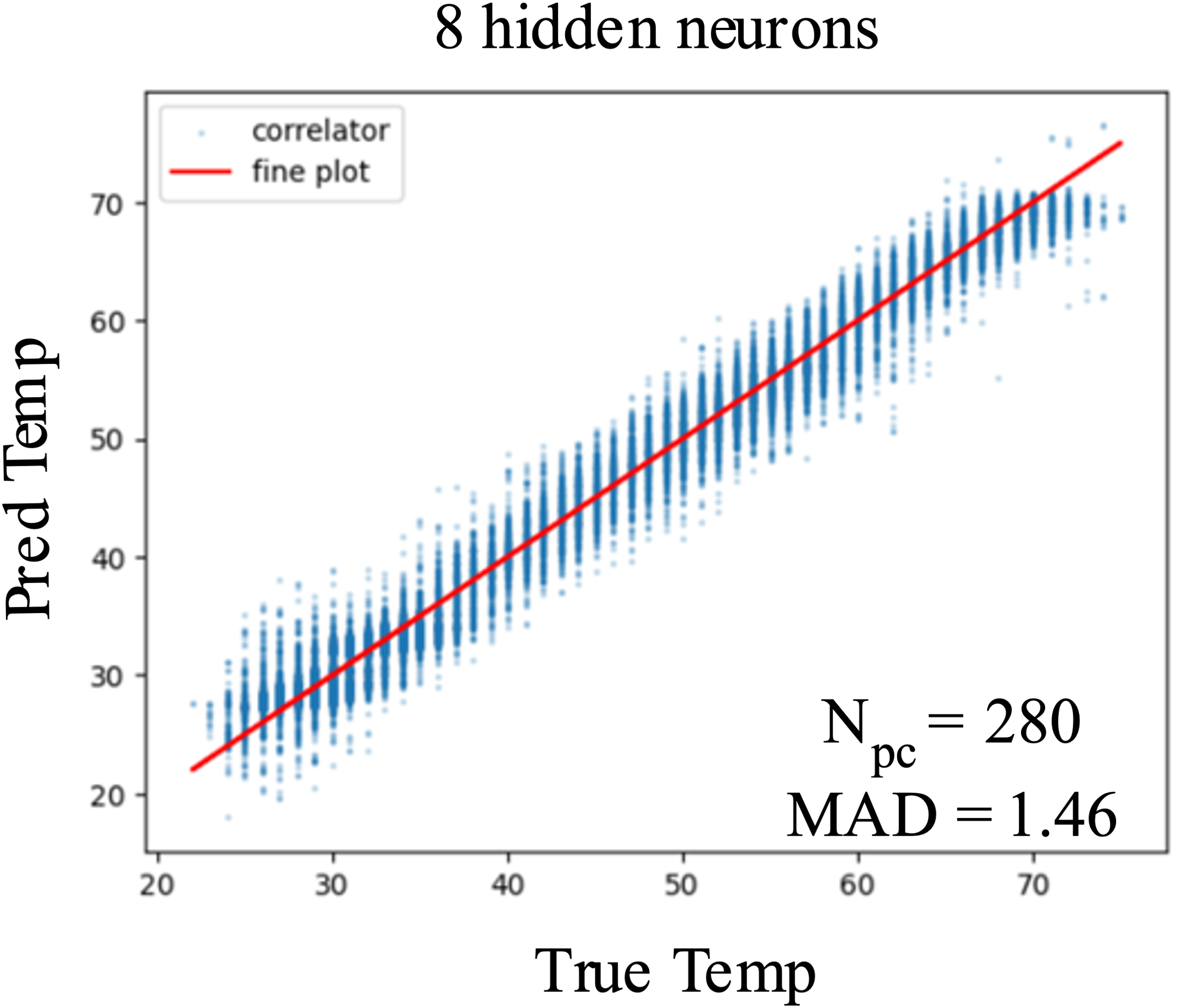}
    \caption{Correlation -- dependence of predicted temperature on true temperature. Fine plot depicts true temperature. Neural network consists of 1 hidden layer with 8 neurons.}
    \label{fig:corr-nn8}
\end{figure}

\begin{figure}[htbp]
    \centering
    \includegraphics[width=0.7\linewidth]{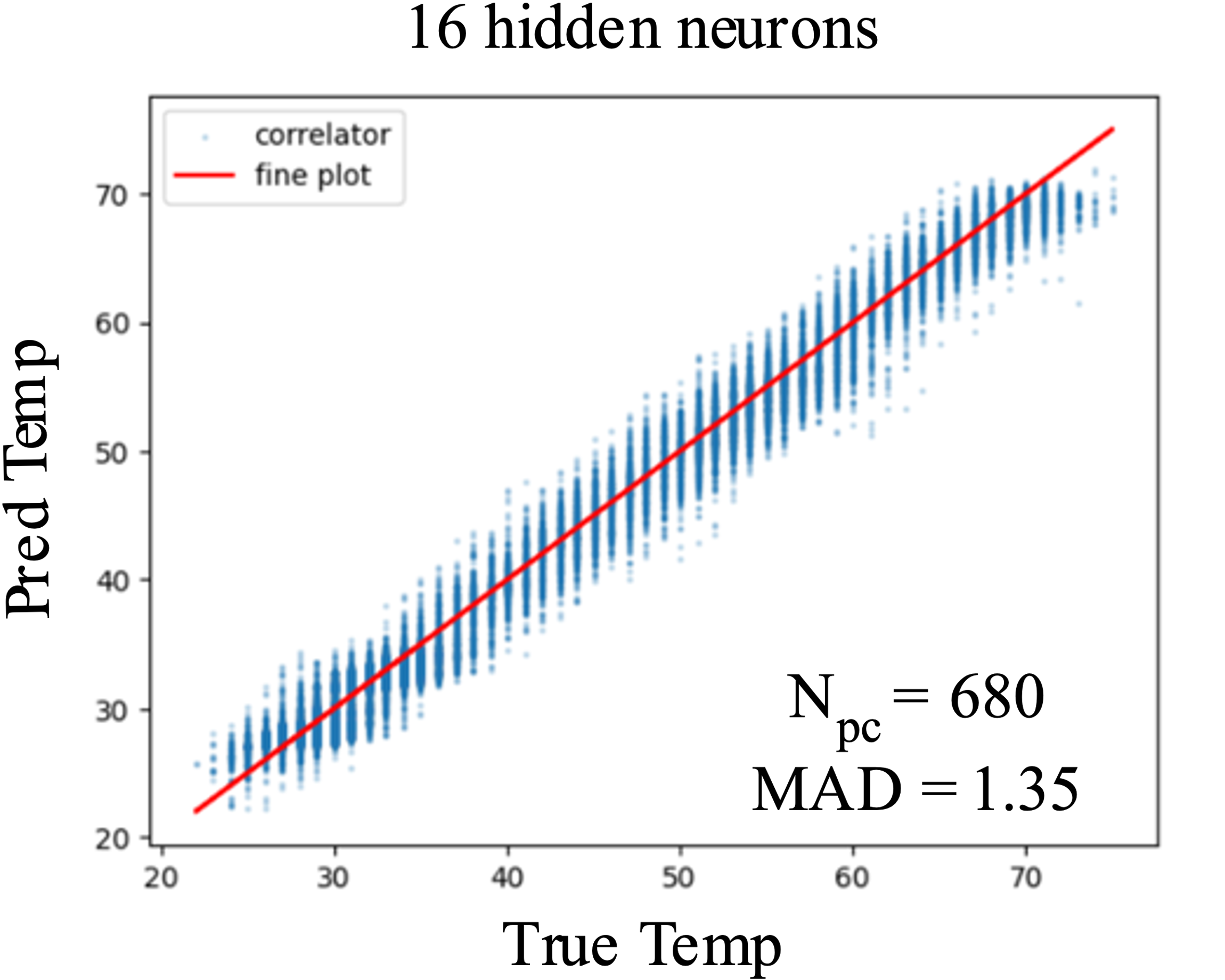}
    \caption{Correlation -- dependence of predicted temperature on true temperature. Fine plot depicts true temperature. Neural network consists of 1 hidden layer with 16 neurons.}
    \label{fig:corr-nn16}
\end{figure}

The next two experiments were conducted using a classical fully-connected neural network. Figure~\ref{fig:corr-nn8} shows the correlator for a neural network with three hidden layers, each layer containing eight neurons. The activation function here was tanh. This system contains 280 parameters. The average deviation was MAD = 1.46~\degC. The maximum deviation was 11~\degC. Finally, Figure~\ref{fig:corr-nn16} shows the best results obtained by a classical neural network containing three hidden layers with 16 neurons. The average deviation was MAD = 1.35~\degC. The maximum deviation was also 11~\degC.

\begin{figure}[htbp]
    \centering
    \includegraphics[width=0.7\linewidth]{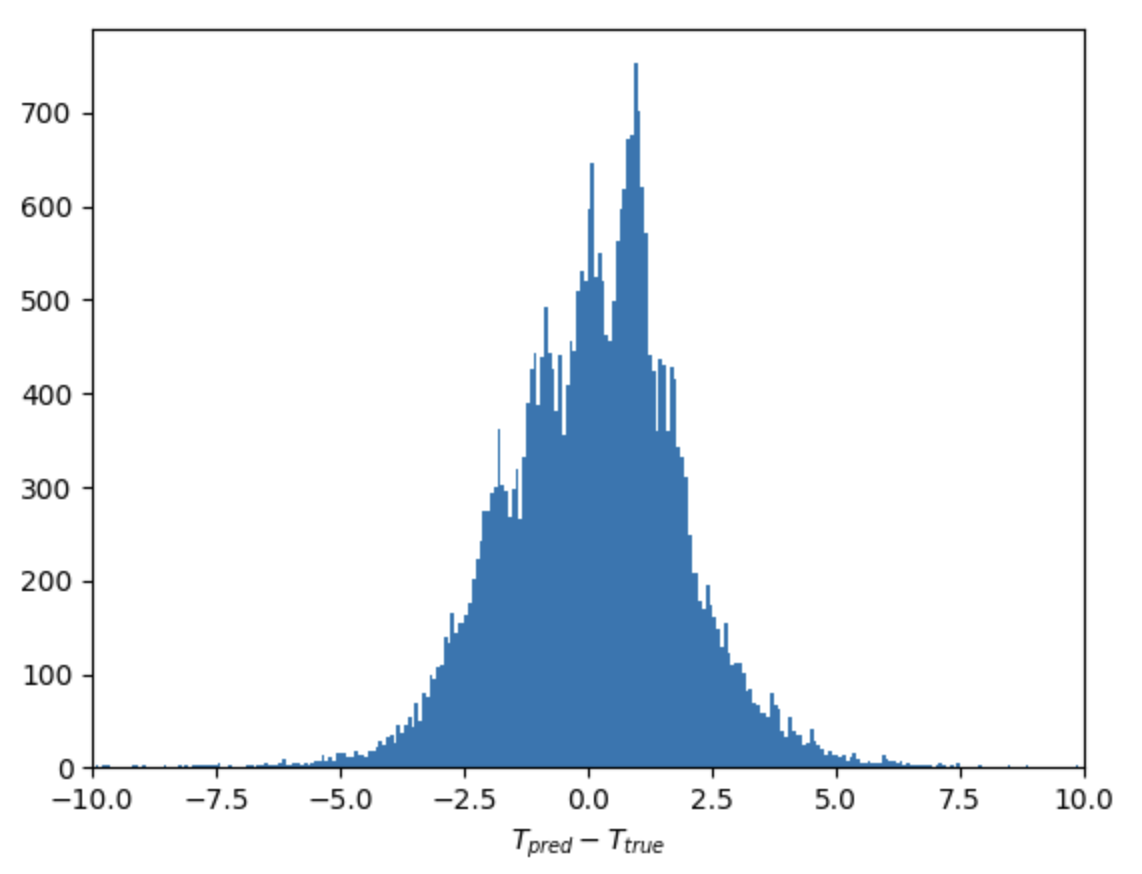}
    \caption{Histogram of difference predicted and true temperatures for quantum circuit consisting of 8 qubits and 12 layers of ansatz.}
    \label{fig:hist-err}
\end{figure}

\section{Discussion}
\label{sec:discussion}

The most important result of the study is the confirmation of the ability of quantum circuits to solve real-world practical problems. Also we obtaine new theoretical results.

The computational experiments confirmed the following patterns:
\begin{itemize}
    \item Quantum machine learning algorithms enable to traine the regression models for predicting of the values of complex functions describing various physical processes,
    \item They enable the modeling of various patterns described by combinations of periodic, decaying, polynomial, and step functions,
    \item They are preferred for modeling periodic and decaying processes. Furthermore, a quantum circuit perfectly approximates a fully harmonic function (e.g., $y = \sin(x)$) -- the mean deviation is MAD = 0.
\end{itemize}

However, quantum circuits with different architectures show different results, and the re-uploading scheme has better performance over a simple quantum circuit. The re-uploading scheme is similar to the residual connection technique in neural networks~\cite{he2016}. Possibly that the quantum circuits have the ability to forget signals like multilayer neural networks. This point should be the subject of further research.

By using an expression metric (the network's ability to approximate complex functions with the fixed number of parameters) for comparing of the performance of networks of different architectures, we observed the effect that predicted by the universal approximation theorem~\cite{cybenko1989}. Specifically, prediction accuracy increases with increasing number of qubits and layers.

Furthermore, the increase of prediction accuracy as result of increasing the number of layers in the quantum circuit confirms that the quantum circuit itself provides the primary contribution to accurate prediction, with the input and output layers acting as feature layers.

Despite of telemetry data noising caused by acting of the server fan control system and thermal hysteresis, the model predicts server temperatures for given power consumption and utilization with accuracy that sufficient for server overheating detection as a result of fans and control system failures.

The average deviation of the predicted temperature was MAD = 1.41\degC\ with maximum deviation around 15\degC. However, the histogram in Figure~\ref{fig:hist-err} shows that an error greater than 5\degC\ is extremely rare. Based on these points we consider a measured temperature exceeding the predicted temperature by 4\degC\ as a signal of hardware failure.

In the next steps, we plan to study the performance of quantum circuits taking into account the quantum noise, as well as examine the temperature prediction capabilities of both approaches depending on the dataset size and the ability of overfitting of the quantum circuit.

A practical problem solution showed that the neural network slightly outperformed the results of the quantum circuit (Figure~\ref{fig:corr-nn16}). This may be due, among other things, to the fact that the server temperature depends on the active influence of the server management system and fan operation, including possible stepwise switching between different operating modes. Further research should clarify the range of conditions under which quantum circuits and quantum machine learning algorithms can effectively implement quantum supremacy.

\section{Conclusions}
\label{sec:conclusions}

In our work, we demonstrated that the quantum circuit is capable of solving the practical problem of server temperature prediction, as well as a range of other regression tasks. All computations were performed on a quantum device simulator. The results show that this approach to solving practical problems can be implemented on a real quantum computer.

\section*{Conflict of Interest}

The authors of this work declare that they have no conflicts of interest.

\bibliography{references}

\end{document}